\documentclass[a4paper,11pt]{article}
\usepackage{aaskaiid}
\usepackage{subcaption} % for subfigures
\usepackage{orcidlink}

\newcommand{\hi}{\textrm{H\textsc{i}}}

\title{Methodological Frontiers in 21-cm Intensity Mapping: the Treatment of Systematics and Foreground Contamination}
\ShortTitle{Methodological frontiers in 21 cm Intensity Mapping}

\ShortName{Spinelli et al.} % shortened name list for header 
\author[1,2]{Marta Spinelli\orcidlink{0000-0003-0148-3254}}
\author[3,8,10,23]{Matilde Barberi-Squarotti\orcidlink{0009-0007-8964-5807}}
\author[4]{José Luis Bernal\orcidlink{0000-0002-0961-4653}}
\author[5]{Somnath Bharadwaj}
\author[6]{Victor Bonjean}
\author[7,2]{Philip Bull\orcidlink{0000-0001-5668-3101}}
\author[8]{Carmelita Carbone}
\author[9]{Isabella P.\ Carucci\orcidlink{0000-0001-5287-0065}} 
\author[10]{Stefano Camera}
\author[2]{Suman Chatterjee}
\author[7]{Tianyue Chen\orcidlink{0000-0003-0173-6274}}
\author[11]{Zhaoting Chen} 
\author[12]{Samir Choudhuri}
\author[13]{Devin Crichton\orcidlink{0000-0003-1204-3035}}
\author[14,7]{Steven Cunnington\orcidlink{0000-0001-6594-107X}}
\author[8]{Bianca De Caro}
\author[12]{Khandakar Md Asif Elahi\orcidlink{0000-0003-1206-8689}} 
\author[15,22,2]{Jose Fonseca\orcidlink{0000-0003-0549-1614}} 
\author[6]{Athanasia Gkogkou\orcidlink{0000-0002-6293-4998}} 
\author[16]{Abhik Ghosh}
\author[7]{Keith Grainge\orcidlink{0000-0002-6780-1406}}
\author[2]{Wenkai Hu}
\author[17,2]{Melis O.\ Irfan\orcidlink{0000-0003-2021-7357}}
\author[18]{Yichao Li}
\author[2]{Siyambonga Matshawule} 
\author[2]{Geoff Murphy}
\author[1]{Simon Prunet\orcidlink{0000-0002-1755-4582}}
\author[2,19]{Mario G.\ Santos\orcidlink{0000-0003-3892-3073}} 
\author[20]{Jean-Luc Starck}
\author[21,2]{{Jingying Wang}\orcidlink{0000-0002-5598-2668}}
\author[7]{Laura Wolz} 
\author[7]{Zheng Zhang\orcidlink{0000-0002-9154-2803}}

\affiliation[1]{Observatoire de la Côte d’Azur, Laboratoire Lagrange, Bd de l’Observatoire, 06304 Nice, France}
\affiliation[2]{Department of Physics \& Astronomy, University of the Western Cape, Cape Town 7535, South Africa}
\affiliation[3]{Dipartimento di Fisica, Universit\`a degli Studi di Milano, via G.\ Celoria 16, 20133 Milano, Italy}
\affiliation[4]{Instituto de Física de Cantabria (IFCA), CSIC-Univ. de Cantabria, Avda. de los Castros s/n, E-39005 Santander, Spain}
\affiliation[5]{Department of Physics, Indian Institute of Technology Kharagpur, Kharagpur 721 302, India}
\affiliation[6]{Institutes of Computer Science and Astrophysics, Foundation for Research and Technology Hellas (FORTH), Greece}
\affiliation[7]{Jodrell Bank Centre for Astrophysics, Department of Physics \& Astronomy, The University of Manchester, Manchester M13 9PL, UK}
\affiliation[8]{INAF, Osservatorio Astrofisico di Brera-Merate, via Brera 28, 20121 Milano, Italy}
\affiliation[9]{INAF, Osservatorio Astronomico di Trieste, Via G.B.\ Tiepolo 11, 34131 Trieste, Italy}
\affiliation[10]{Dipartimento di Fisica, Universit\`a degli Studi di Torino, via P.\ Giuria 1, 10125 Torino, Italy}
\affiliation[11]{Institute for Astronomy, The University of Edinburgh, Royal Observatory, Edinburgh EH9 3HJ, UK}
\affiliation[12]{Centre for Strings, Gravitation and Cosmology, Department of Physics, Indian Institute of Technology Madras, Chennai 600036, India}
\affiliation[13]{Institute of Particle Physics \& Astrophysics, Department of Physics, ETH Zurich, 8093 Zurich, Switzerland}
\affiliation[14]{Institute of Cosmology \& Gravitation, University of Portsmouth, Dennis Sciama Building, Portsmouth, PO1 3FX, UK}
\affiliation[15]{Instituto de Astrof\'isica e Ci\^encias do Espa\c{c}o, Universidade do Porto CAUP,Rua das Estrelas, 4150-762 Porto, Portugal}
\affiliation[16]{Department of Physics, Banwarilal Bhalotia College, Asansol, West Bengal-713303, India}
\affiliation[17]{Institute of Astrophysics, University of Cambridge, Madingley Road, CB3 0HA}
\affiliation[18]{Department of Physics, College of Sciences, Northeastern University, Wenhua Road, Shenyang, China}
\affiliation[19]{South African Radio Astronomy Observatory (SARAO), Cape Town, 7700, South Africa}
\affiliation[20]{Universit\`e Paris-Saclay, Universit\`e Paris Cit\`e, CEA, CNRS, AIM, 91191, Gif-sur-Yvette, France}
\affiliation[21]{Shanghai Astronomical Observatory, Chinese Academy of Sciences, 80 Nandan Road, Shanghai 200030, China}
\affiliation[22]{Departamento de F\'isica e Astronomia, Faculdade de Ci\^{e}ncias, Universidade do Porto, Rua do Campo Alegre 687, 4169-007 Porto, Portugal}
\affiliation[23]{INFN -- Istituto Nazionale di Fisica Nucleare, Sezione di Milano, via G.\ Celoria 16, 20133 Milano, Italy}
\emailAdd{marta.spinelli@oca.eu}

\abstract{
The distribution of neutral hydrogen (\hi) in the post-reionization universe traces the cosmic large-scale structure and therefore serves as a powerful cosmological probe. An efficient way to measure its distribution over large sky areas and wide redshift ranges is through single-dish intensity mapping, which exploits the autocorrelation signal of each dish in a telescope array, while they are scanning the same patch in the sky. Thanks to its broad frequency coverage and technical capabilities, SKA-Mid will enable measurements of the integrated 21 cm emission from \hi\ up to redshift $z\sim3$, making single-dish intensity mapping a key observable for probing dark matter and dark energy.

\medskip
Isolating the faint 21 cm cosmological signal without introducing distortions or biases is, however, a challenging task. The 21 cm signal is several orders of magnitude weaker than the astrophysical foregrounds and its analysis is further affected by instrumental systematics. Overcoming these difficulties requires detailed modelling, as well as continuous improvements and innovations in data analysis techniques. Over the past decade, the international community has been developing and testing new methods, both to address current observational challenges and to prepare for the forthcoming SKA-Mid observations.

\medskip
This chapter presents a review of recent advances in map-making and component separation techniques, with particular emphasis on dealing with telescope-specific systematics such as beam response and correlated noise. We focus on results obtained in controlled simulation environments, providing a valuable framework for assessing the strengths and limitations of different approaches.

\medskip
Developing robust algorithms capable of accurately handling instrumental effects and sky-model uncertainties is a crucial step toward fully exploiting the cosmological potential of \hi\  intensity mapping surveys in the SKA Observatory era.

}

\begin{document}
\maketitle

\section{Introduction}
Intensity Mapping (IM) is a technique to trace the cosmic distribution of neutral hydrogen directly from the integrated luminosity of its redshifted 21 cm line, without the need for the detection of individual sources. 
Despite the sensitivity performance of the SKA Observatory (SKAO) \citep{braun2019anticipatedperformancesquarekilometre}, the detection of neutral hydrogen clouds beyond the local Universe will
remain a challenge, and only a small number of very bright galaxies at 21 cm will be detected at $z\sim1$. IM will instead significantly extend the redshift range accessible for observing the large-scale distribution of neutral hydrogen, thus accessing the underlying matter distribution as a function of cosmic time \citep[e.g.][]{wyitheloeb2008, bagla2010, Ansari2012, Battye-2013}.

With the SKA-Mid precursor MeerKAT, the MeerKLASS collaboration has pioneered the approach of performing IM operating the entire array as a fast-scanning collection of multiple individual dishes and exploiting the autocorrelation signals, also known as “single-dish” mode \citep{santos2017}. This enabled the mapping of the sky on angular scales that the interferometer would resolve out, while using one of MeerKAT's strengths: the speed and precision of the antennas' synchronized movement. A summary of the work and results achieved via single-dish IM on MeerKAT in recent years is presented in chapter \citet{Cunnington01.2026.SKA}. These results show that the technique, originally proposed as an idea in the first edition of this book about 10 years ago \citep{santos2015}, has evolved into a concrete reality. The experiences gained with MeerKAT, including lessons learned, challenges overcome, and best practices developed, provide a foundation for SKA-Mid to achieve even more robust results, marking a significant step toward the full maturity of radio cosmology.

Indeed, a single-dish survey with SKA-Mid for the post-reionization universe will allow us to place tight constraints on cosmological parameters that are both competitive and complementary to current galaxy surveys, challenging our understanding of dark matter and dark energy. A detailed description of the anticipated impact on cosmology of an intensity mapping survey  with SKA-Mid is described in detail in \citet{Wolz01.2026.SKA,Camera01.2026.SKA, Fonseca01.2026.SKA}, where the interested reader can also find a carefully compiled list of references. A summary of the methodologies developed to capture and exploit non-Gaussian features in 21 cm signals from future SKA-Mid observations via various higher-order statistics can instead be found in \citet{Majumdar01.2026.SKA}.

The process of transforming the autocorrelation signals from each SKA-Mid antenna observing the same patch of sky, into a single 21 cm data cube that accurately represents the cosmic 21 cm signal and is ready for cosmological interpretation, will not be trivial. It will require constant attention to the response and instrumental characteristics of the telescope, as well as continuous improvements and innovations in analysis techniques to handle contaminants and systematic effects. The interested reader can find a detailed description of the Radio Frequency Interference (RFI) flagging, calibration, and map-making pipeline currently developed for MeerKAT in \citet{Wang2021} and \citet{MeerKLASS2025}.

Another critical challenge, coupled with calibration and map-making, lies in isolating the faint 21 cm signal without distorting or biasing its intrinsic cosmological information. The cosmological \hi\ emission is buried beneath much brighter foreground radio emissions that dominate the same frequency range. The strongest of these, synchrotron emission from our own Galaxy, exceeds the expected 21 cm signal by roughly five orders of magnitude, even at high Galactic latitudes. Similarly, extragalactic point sources can be about three orders of magnitude brighter. Such strong foregrounds can severely contaminate the data, unless they are removed with exceptional precision.

In the last decade, many studies have addressed the problem in the context of single-dish \hi\ IM and investigated the quality of foreground removal methods on the data \citep{Switzer:2015ria, Wolz:2015lwa, Cunnington:2020wdu,mPCA}, as well as simulations \citep[e.g.,][]{Ansari2012, Wolz:2013wna, Alonso-2014, Shaw2015, Olivari-2016, Carucci:2020enz,  mlv+21, 2021MNRAS.504.5231Y, 2021MNRAS.504..267F, Soares2021,Spinelli2021}, where blind and non-parametric methods have proven to be the most powerful. 

A further complication arises from the instrument itself, which can introduce additional systematics such as the telescope beam response, amplitude gain fluctuations, polarization leakage, internal signal-chain reflections, and digital non-linearities. Overcoming or at least mitigating these effects requires meticulous calibration and a deep understanding of instrumental behaviour. 

The telescope’s primary beam, for example, affects both calibration and component separation, imprinting its own frequency dependence on the foregrounds. Accurately characterizing the beam, particularly far from its centre, is difficult due to variations between individual dishes and temporal changes. Understanding the extent to which these instrumental effects may hinder the detection of the cosmological signal is therefore of paramount importance.

The controlled framework of simulations allows the investigation of these individual observational challenges of \hi\ IM. Studies related to SKAO, its precursors, and pathfinders include the primary beam effects \citep{Matshawule2021,Spinelli2021,Gkogkou2026}, polarization leakage \citep{Shaw2015, Carucci:2020enz, Cunnington:2020njn}, $1/f$ noise \citep{Harper181f, Chen20201f, li21, irfan24} and radio frequency interference due to satellites \citep{Harper181f,Engelbrecht2025}. 
%The necessity of a realistic instrument characterization and survey design makes end-to-end simulations a crucial requirement towards a valid detection of the \hi\ IM signal in auto-correlation.

%This chapter provides a review of recent techniques developed both to pave the way for the SKAO era and to tackle the immediate challenges that SKAO precursors and pathfinders are facing in analysing their data. These novel methodologies are mostly presented in the context of simulations where it is possible to mimic specific observational effects. The discussion is necessarily selective, and readers seeking further details are encouraged to consult the references provided. The list is also not exhaustive and predominantly reflects contributions from the Cosmology Science Working Group (SWG).

The chapter is organized as follows. In \autoref{sec:sim}, we provide a very brief overview of the typical choices for the simulated single-dish IM observations. We present a brief review of the problem of component separation in the context of intensity mapping highlighting the difference among blind and non-blind methods in \autoref{sec:FG_intro}. In \autoref{sec:FG_blind}, we focus on blind component separation techniques providing a brief review of the various methodologies discussed in literature and typically developed for solving a similar problem in the context of the Cosmic Microwave Background (CMB). We summarize the finding of a component separation data challenge that we performed within the Cosmology Science Working Group (SWG) in \autoref{sec:challenge}.
In \autoref{sec:beam}, we deal more specifically with the added complexity of a primary beam sidelobes and frequency evolution. In \autoref{sec:news} we explore instead component separation methods that relies on strategies different from blind methods, including the the multi-frequency angular power spectrum approach, AI driven solutions and treating the problem in higher dimensional space using Bayesian statistics. 
In \autoref{sec:loss}, we discuss the issue of signal loss in component separation. In \autoref{sec:pink} we focus on the $1/f$, on how this can be characterized and dealt with at map-making process. We conclude in \autoref{sec:moresys} where we provide a brief final review on other possible types of systematics and the challenges that the community will need to tackle for to fully exploit the capabilities of SKA-Mid for precision large scale structure studies with \hi\ IM. 

\section{Typical choices for IM simulations}\label{sec:sim}
A typical IM mock data cube contains: i) foreground emission that are bright in the frequency range of interest (SKA-Mid Band 1 and 2), ii) simulated temperature brightness from the 21 cm line at the corresponding redshifts (up to $z\sim 3$ for Band 1),  iii) a simplified description of the effect of the telescope beam, and iv) a white noise realisation from the radiometer equation. Since the noise level decreases with the square root of the number of dishes, the SKA-Mid AA4 configuration, with 197 dishes, is expected to achieve a lower noise level than the AA* configuration, which includes 144 dishes. Therefore, AA4 represents our ideal-case scenario.

The astrophysical foregrounds often included in the mock data are galactic synchrotron and free-free emissions, and extragalactic radio point sources. The synchrotron and free-free emissions are generally modelled as smooth power laws in frequency, with amplitudes and spectral indices provided by the Planck Legacy Archive FFP10 simulations or the Planck Sky Model \citep{delabrouille2013} (in particular via the popular \textit{pysm} software \citep{Thorne_2017}).  For extragalactic radio point sources (i.e., active galactic nuclei, radio galaxies, quasars, etc.), empirical models \citep[e.g.][]{Battye-2013} or information from available surveys \citep[e.g.][]{Matshawule2021} are used. Simulated leakage of polarized foreground to total intensity has been also considered often using the models from \citet{Alonso-2014,Shaw2015}. The polarised signal could have frequency structure due to Faraday rotation \citep{spinelli2018} complicating the IM analysis. 

The cosmological \hi\ maps are normally computed using lognormal realisation of the matter field and then transformed into temperature brightness assuming a scale-independent \hi\ bias. A popular software is described in \citet{Alonso-2014}. With the progress of the data quality and the data analysis tools in the last decade, the need for more realistic simulations that could help the interpretation of the signal has become more pressing. A review of some of the available simulations that are used by the IM community can be found in \citet{Ronconi01.2026.SKA}. Simulations including more sophisticated \hi\ treatment and galaxy properties are useful in particular for cross-correlation analysis with galaxy surveys \citep[e.g.][]{Wolz:2021ofa}. 

To mimic an SKA-Mid observation, the maps are often smoothed with a frequency-dependent Gaussian telescope beam (which we will discuss in more detail in \autoref{sec:beam}) and add Gaussian thermal noise, following the prescriptions of \citet{SKA-2018}. 

%\begin{table}[]
%\centering
%\begin{tabular}{lll}
%\hline
%\multicolumn{3}{c}{SKAO-MID AA4 specifications}                  \\ %%\hline
%Number of dishes             & $N_{\rm dish}$ & 197       \\
%Diameter of the dishes       & $D_{\rm dish}$ & 14.51 m   \\
%Observed fraction of the sky & $f_{\rm sky}$    & 50\%      \\
%Observation time             & $t_{\rm obs}$    & 10\,000 hrs \\
%Spill-over temperature       & $T_{\rm spill}$  & 3 K       \\
%Channel width                   & $\Delta \nu$     & 1 MHz     \\
%\hline
%\end{tabular}
%\caption{Instrumental specifications for computing the thermal noise and the telescope beam. Note that the number of dishes includes both the SKAO and the MeerKAT dishes.}
%\label{tab:SKA_spec}
%\end{table}

%The data cube consists of a set of 105 \textsc{HEALPix} (\cite{Gorski-2004}) maps, with resolution $N_{\rm side}=128$ and thickness of 1 MHz. The frequency range is $\nu \in [900.5,\,1004.5]$ MHz, corresponding to the redshift range $z \in [0.41,\,0.58]$. In this range, the beam width varies from 1.3 to 1.8 deg. All simulated maps are masked in the Galactic plane, with $f_{\rm sky}=50\%$.

\section{Foreground cleaning methodology}\label{sec:FG_intro}

There are a variety of strategies to deal with these extragalactic and Galactic foreground contaminations. Broadly speaking, one can categorize them into a spectrum of methods going from foreground avoidance, to blind foreground subtraction, to semi-blind or non-blind methods, where some level of knowledge on the components is imposed in the solution. 

\emph{Foreground avoidance} finds an observational window at the power spectrum level in $k_{||} - k_\perp$ space, where foregrounds do not substantially affect the estimates of the power spectrum \citep[e.g.]{Shaw2015}. \emph{Blind foreground subtraction} methods, that are described in more detail in \autoref{sec:FG_blind}, have been the most successful in the last years, especially facing real data.
There are also \emph{methods involving some knowledge of the foregrounds}. Some earlier attempts to subtract foregrounds imposed a polynomial logarithmic structure in frequency \citep{2009ApJ...695..183B}. Others have used MeerKLASS data to calibrate the spectral index of galactic synchrotron \citep{2022MNRAS.509.4923I} and gain insight of the foregrounds which can later be used on foreground removal. In \citet{2021MNRAS.504..267F}, they used the models of \citet{2005ApJ...625..575S} to study if one could marginalize over the foregrounds models and learn about their angular structure. These parametric approaches are not the sole methods using prior knowledge of contaminants: the authors of \citet{2024MNRAS.534.2653M} model some of the instrumental effects which are then marginalized over using an Hamiltonian Monte Carlo sampler.

%%%%%%%%%%%%%%%%%%%%%%%%%%%%%%%%%%%%%%%%%%%%%%%%%%%%%%%%%%%%%%%
\section{Blind component separation} \label{sec:FG_blind}

Several widely used and well-tested methods for addressing the challenge of separating contaminants in 21 cm IM rely only on two main assumptions: i) foreground emissions vary slowly with frequency compared to the rapidly oscillating cosmological signal and, ii) foreground emission is significantly larger in magnitude than the \hi\ signal. We can identify the primary foreground contributions as a set of smooth functions of frequency, which can be subtracted from the data to isolate the \hi\ signal. 

In practice, IM observations can be represented by two-dimensional (frequency and pixel) data-cubes, $\boldsymbol{\mathsf{X}} \equiv  T_{\rm obs}(\nu,\hat{\textbf{n}})$. It is then assumed that the matrix $\boldsymbol{\mathsf{X}}$, dominated by the foreground emission, can be linearly decomposed into a set of $N_{\rm fg}$ ``foreground" sources in pixel space $\boldsymbol{\mathsf{S}}$ modulated in frequency through a mixing matrix $\boldsymbol{\mathsf{A}}$ plus some residuals $\boldsymbol{\mathsf{N}}$ that should in principle contain most of the cosmological signal that we aim to recover plus the white instrumental noise: %  
\begin{equation}\label{eq:systosolve}
    \boldsymbol{\mathsf{X}}=\boldsymbol{\mathsf{A}}\,\boldsymbol{\mathsf{S}} + \boldsymbol{\mathsf{N}} \,.
\end{equation}

The process of contaminant separation involves solving \autoref{eq:systosolve} to determine $\boldsymbol{\mathsf{N}}$. The assumptions made to derive matrix $\boldsymbol{\mathsf{A}}$ and components $\boldsymbol{\mathsf{S}}$ that satisfy \autoref{eq:systosolve} differ among various methods. 
In the next sections, we outline the main ideas following \citet{Spinelli2021}. We refer interested readers to the various references discussed below for a more accurate treatment of the problem.

\subsection{Principal Component Analysis}
\label{ss:pca}

Principal Component Analysis (PCA) can be used to identify an estimate for the mixing matrix $\boldsymbol{\mathsf{A}}$, the columns of which will be given by the first $N_\text{fg}$ principal components. The principal components are essentially the eigenvectors of the mean-centred data $\nu\nu'$ covariance matrix $\boldsymbol{\mathsf{C}}$, given by

\begin{equation}\label{eq:InverseNoiseCov}
    C_{ij}=\frac1{N_{\hat n}}\sum^{N_{\hat n}}_{p=1} w_i \Delta T(\nu_i,\hat n_p)~ w_j \Delta T(\nu_j,\hat n_p)\,,
\end{equation}

where $\Delta T(\nu_i,\hat n_p) = T(\nu_i,\hat n_p)-\bar T(\nu_i)$ and the summation is over all $N_{\hat n}$ pixels . The $w$ factors provide an optional map weighting.
The eigendecomposition is given by $\boldsymbol{\mathsf{C}}\,\boldsymbol{\mathsf{V}} = \boldsymbol{\mathsf{V}}\,\mathbf{\Lambda}$, where $\mathbf{\Lambda}$ is the diagonal matrix of $N_\nu$ eigenvalues. The first $N_\text{fg}$ columns from the eigenvector matrix $\boldsymbol{\mathsf{V}}$ represent the entries for the mixing matrix.

\subsubsection{Multiscale PCA}
It has recently been proposed to take advantage of PCA in a multiscale framework to complement with angular information its frequency-direction-only assumptions on the contaminants \citep{mPCA}. Each frequency two-dimensional map is first decomposed into wavelet-filtered large- and small-scale maps, using the isotropic undecimated wavelet transform \citep{Starck2007}. The original data cube now corresponds to two cubes, referring either to the large- or small-scale only information, which are then cleaned separately. The component separation process becomes two independent processes, with their own, independently defined mixing matrix and number of removed components, $N_\text{fg}$, resulting in a better characterization of the contaminants. The final cleaned map is assembled by summing the two independently cleaned small- and large-scale maps, since the wavelet decomposition is numerically exact.

\subsection{Independent Component Analysis}

Independent Component Analysis (ICA) is a blind source separation method that extracts statistically independent signals without prior knowledge of how they were combined. It assumes that the data are linear combinations of independent sources and  identify them  by maximizing any statistical quantity that measures non-Gaussianity.
In particular, Fast Independent Component Analysis (FastICA), developed in \citet{Hyvrinen1999FastAR}, has been employed for foreground cleaning on simulated \hi\ data \citep{Chapman:2012yj,Wolz:2013wna, Alonso-2014,Cunnington:2019lvb,Carucci:2020enz,Cunnington:2020njn} as well as real data \citep{Wolz:2015lwa,Wolz:2021ofa, Hothi2021}. 
Since it often contains a ``whitening'' step that can be achieved with a PCA analysis, FastICA is essentially an extension of PCA, and hence in most cases in the context of foreground cleaning, will provide very similar results.

\subsection{Generalised Morphological Component Analysis}
\label{sec:gmca}

Generalised Morphological Component Analysis (GMCA) is a blind component separation method based on sparsity \citep{gmca1} that has been successfully applied in various astrophysical contexts. It assumes that the $N_\text{fg}$ foreground components verify two hypotheses: they are sparse in a given transformed domain (i.e., most samples are zero-valued) and their supports are disjoint; in other words, the foreground components are {\it morphologically} diverse (i.e., their non-zero samples appear at different locations). 

\citet{Carucci:2020enz} showed the wavelet domain to be optimal to sparsely describe foregrounds and contaminants in the low-$z$ \hi\ IM context. Firstly, we project the data $\boldsymbol{\mathsf{X}}$ onto wavelet space. The GMCA algorithm aims at minimizing the following cost function:
\begin{equation} \label{eq:GMCAmaster}
   \min_{\boldsymbol{\mathsf{A}}, \boldsymbol{\mathsf{S}}}  \sum_{i=1} ^{N_\text{fg}}  \lambda_i \left\lvert \left\lvert S_{i} \right\rvert \right\rvert_1+ \left\lvert \left\lvert \boldsymbol{\mathsf{X}} - \boldsymbol{\mathsf{A}}\,\boldsymbol{\mathsf{S}} \right\rvert \right\rvert_{2}, 
\end{equation}
where the first term is the $\ell_1$ norm, i.e. $\sum_{j,k} \left\lvert  S_{j,k} \right\rvert$: this constitutes a constraint for sparsity, mediated by the regularization coefficients $\lambda_i$. The second term is the usual data-fidelity $\ell_2$ norm term. We find solutions for $\boldsymbol{\mathsf{A}}$ and $\boldsymbol{\mathsf{S}}$ by iterating a projected alternate least-square procedure: we fix $\boldsymbol{\mathsf{A}}$ and perform a least-square update to determine $\boldsymbol{\mathsf{S}}$, we compute the thresholds $\lambda_i$ via mean absolute deviation of $S_i$, we update $\boldsymbol{\mathsf{A}}$ with $\boldsymbol{\mathsf{S}}$ fixed and so on. The key point is the thresholding: it makes possible to keep the samples with the highest amplitudes, which are the most informative to retrieve the mixing matrix $\boldsymbol{\mathsf{A}}$ (i.e., they most likely to belong to the foreground components and are the least likely to be contaminated by the cosmological signal and noise), and it provides robustness in terms of convergence since the thresholds decrease with the progressive iterations.

%%%%%%%%%%%%%%%%%%%%%%%%%%%%%%%%%%%%%%%%%%%%%%%%%%%%%%%%%%%%%%%%%%%
\subsection{Generalized Needlet Internal Linear Combination} \label{ss:gnilc}

The Internal Linear Combination (ILC) technique was initially introduced in the analysis of WMAP CMB data \citep{Bennett-2003}. It extracts one or more components with known spectral behavior from multi-frequency observations by applying a weight vector $W$. These weights are designed to preserve unit response to the target signal while minimizing the overall variance contributed by residual foregrounds.

Lookng back at \autoref{eq:systosolve}, the ILC estimate of $\boldsymbol{\mathsf{S}}$ is obtained as a weighted linear combination, $\hat{\boldsymbol{\mathsf{S}}} = \boldsymbol{\mathsf{W}}^{\sf T}\,\boldsymbol{\mathsf{X}}$. A major advantage of the ILC approach is that it does not rely on explicit modeling of the foregrounds. Its main limitation, however, is that it can only recover components with a spatially invariant frequency dependence.

The Generalized Needlet ILC (GNILC) method extends the ILC framework to recover emissions that cannot be represented by a single template with fixed frequency scaling, such as diffuse foregrounds. It operates in needlet space and incorporates \textit{prior} information on the power spectrum of the target signal to compensate for the lack of precise knowledge of its spectral dependence \citep{Remazeilles-2011}. Needlets are a kind of spherical wavelets introduced for statistical analysis by \citet{Narcowich-2006};  their properties are discussed in \citet{Baldi-2006} and \citet{Marinucci-2011}. Their first application in cosmology was for CMB data analysis \citep{Pietrobon-2006,Marinucci-2008}. The needlet decomposition provides simultaneous localization in both sky position and angular scale, enabling a more flexible component separation.

GNILC has been successfully applied to CMB analysis by \citet{Planck-inter-2016}, and subsequently adapted for 21 cm intensity mapping studies by \citet{Olivari-2016} and the BINGO collaboration \citep{Fornazier-2022}.

The GNILC algorithm requires two main inputs: the theoretical 21 cm power spectrum across frequency channels (in this sense it can be considered a \textit{semi-blind} method) and an estimate of the number of foreground components, $N_{\rm fg}$, which is needed to run a constrained PCA step to separate the \hi\ and the foreground plus noise components. By transforming  the covariance matrix of the observations using the information from the theoretical prior on the HI signal, the constrained PCA identifies the \hi\ subspace as the eigenvectors whose eigenvalues are approximately equal to 1, while the remaining $N_{\rm fg}$  with larger eigenvalues define the foreground plus noise subspace.

This procedure is carried out independently at each needlet scale using the Akaike Information Criterion (AIC) \citep{Akaike-1974}. For each sky location and needlet scale, $N_{\rm fg}$ is determined by minimizing the AIC:
\begin{equation}
    \label{eq:AIC}
    {\rm AIC}(N_{\rm fg}) = 2\,n\,N_{\rm fg} - 2\,\ln{\mathcal{L}_{max}(N_{\rm fg})}\,,
\end{equation}
where $n$ is the number of modes within the needlet domain, and $\mathcal{L}_{\rm max}$ denotes the maximum-likelihood estimate of the data covariance matrix under a model with $N_{\rm fg}$ independent foreground components (see Appendix A of \citet{Olivari-2016} for details). Once $N_{\rm fg}$ has been determined, the ILC weight matrix $\boldsymbol{\mathsf{W}}$ is computed at each needlet scale, allowing reconstruction of the \hi\ signal from the observational data.

\medskip

In \citet{DeCaro-2025}, the GNILC method was tested on simulated \hi\ intensity maps for SKAO-Mid in the AA4 configuration. The assumptions for such simulations are taken from \citet{Carucci:2020enz}, and are typical for most of the works discussed in this review chapter (see \autoref{sec:sim}). The telescope beam is assumed Gaussian, with $\theta_{\rm FWHM}=\frac{c}{\nu D}$, where \textit{D} is the dish diameter, and it varies across the frequency channels.

\begin{figure}[ht]
\centering
\setlength{\tabcolsep}{0.01pt}
\begin{tabular}{cc}
    \includegraphics[width=0.49\linewidth]{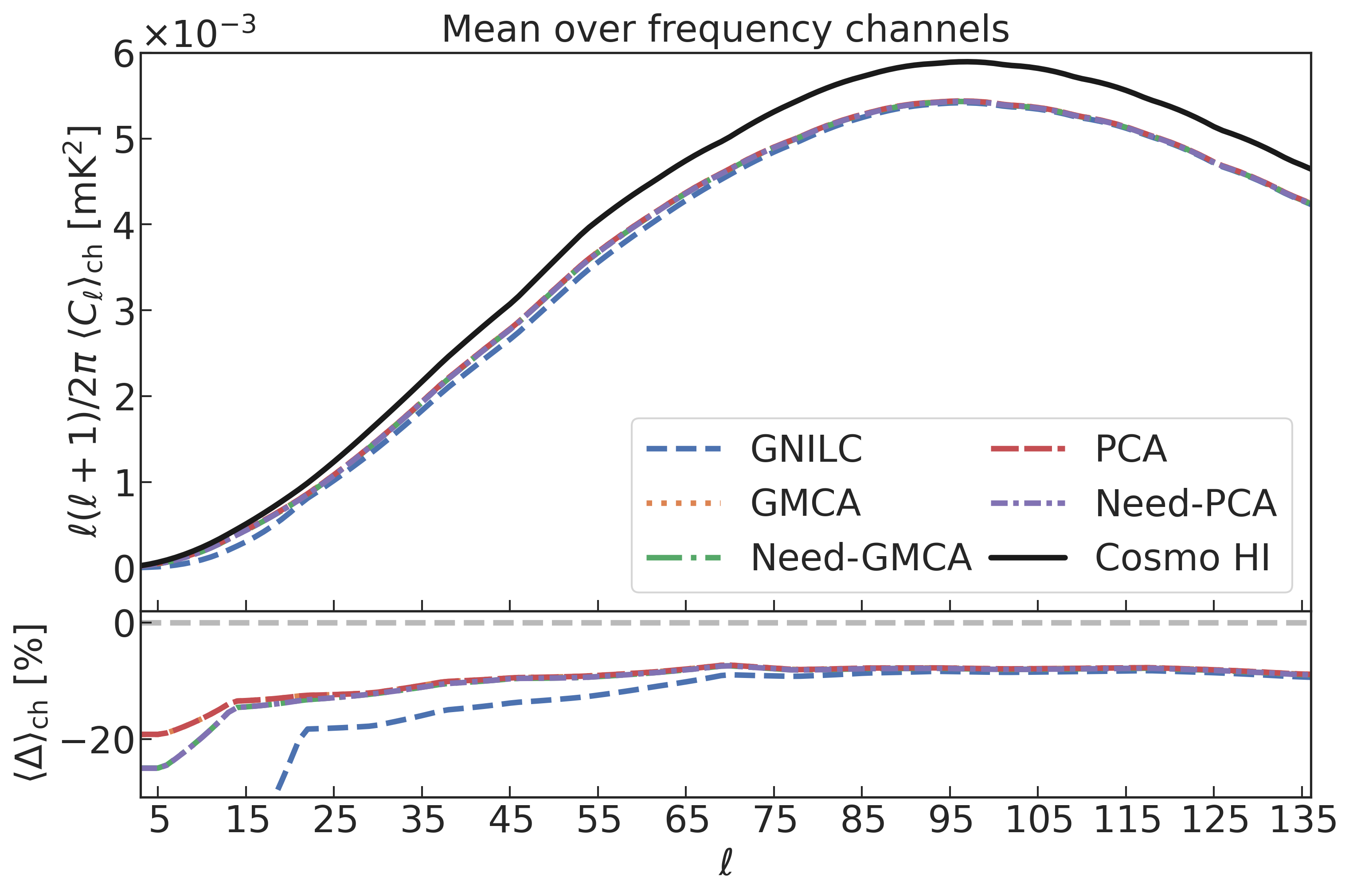}
    \includegraphics[width=0.49\linewidth]{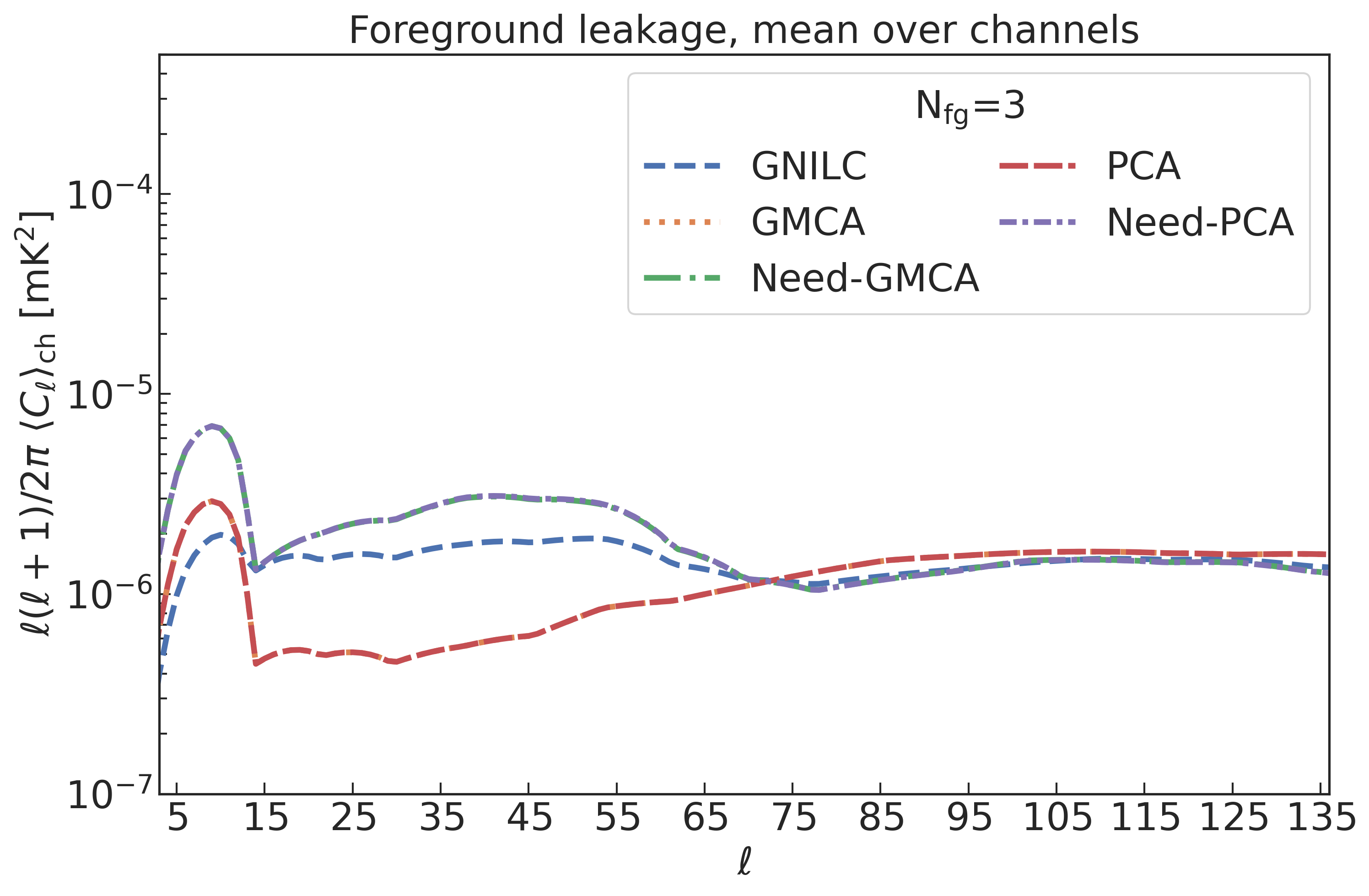}
\end{tabular}
    \caption{\emph{Left}. Comparison between the $C_{\ell}^{\rm Res}$ recovered with GNILC (blue line), GMCA (orange line), Need-GMCA (green line), PCA (red line) and Need-PCA (purple line). The input $C_{\ell}^{\rm cosmo}$ is plotted in black. The bottom panel shows the percent difference $\Delta$. The telescope beam is Gaussian, with $\theta_{\rm FWHM}=\frac{c}{\nu D}$, where \textit{D} is the dish diameter, and it varies across the frequency channels. $N_{\rm fg}=$3 for PCA, Need-PCA, GMCA and Need-GMCA; GNILC computes $N_{\rm fg}$ for each scales and region of pixel according to the AIC. \emph{Right}. comparison between the $C_{\ell}$ of the foreground leakage obtained by averaging over the frequency channels. Figures taken from \citet{DeCaro-2025}.}
        \label{fig:panel_summary_SKA_beam}
\end{figure}

They also implement a needlet version of PCA (Need-PCA) and GMCA (Need-GMCA). Each PCA or GMCA step is applied to the needlet coefficients of the observation map after needlet decomposition. The number of foreground components to remove is set at the same $N_{\rm fg}$, for each needlet coefficient. In the left panel of \autoref{fig:panel_summary_SKA_beam}, GNILC performance is compared with PCA and GMCA, both in their standard version and when using needlets, plotting the relative difference between the input \hi\ angular power spectrum and the recovered one. At larger scales ($\ell \lesssim 30$), GNILC underestimates the input power spectrum. At those scales, the cosmological signal and the foregrounds are degenerate; hence GNILC tends to remove more components, leading to smaller foreground residuals in the \hi\ plus noise maps compared to the other methods. Conversely, a fraction of the \hi\ plus noise is also removed along with the foregrounds, causing signal loss.

Projecting the mixing matrix onto the pure-foreground \citep[e.g.][]{Cunnington:2020njn} one  can estimate the foreground that leaks into the cosmological signal plus noise, as shown in the right panel of \autoref{fig:panel_summary_SKA_beam}.
For the very large scales ($\ell<15$), GNILC achieves lower foreground residuals than the other methods, as it dynamically adapts the number of foreground components depending on scale and
sky region, whereas this is fixed in the other methods.
As the galactic plane is excluded in the analysis, considering the polarization leakage following the procedure of \citet{Alonso-2014} does not alter the conclusion. The same is true considering or not a frequency evolving beam (the standard $\lambda/D$ dependency). 

\section{The foreground cleaning data challenge of the Intensity Mapping Focus Group}\label{sec:challenge}

In \citet{Spinelli2021}, a detailed study of foreground removal methods is presented in particular for AA4 configuration SKA-Mid \hi\ IM surveys. The work has been an initiative of the \hi\ Intensity Mapping Focus Group of the SKA Cosmology Science Working Group: a data challenge for component separation where participants were presented with simulated data-cubes of unknown foregrounds, \hi\ signal and instrumental specifics such as the beam and noise level.

Data challenges are a useful test for the maturity of analysis pipelines and have been routinely exploited by the SKA Observatory in recent years \citep{bonaldi2018squarekilometrearrayscience, Bonaldi_2020}. 

The realistic simulations for the task were constructed with two different implementations of the astrophysical foregrounds based on \citet{2005ApJ...625..575S} and the Planck Sky Model \citep{Planck-inter-2016}, with a realistic scanning strategy for a $\sim5000\,\deg^2$ survey resulting in anisotropic noise, as well as a more sophisticated beam model (i.e. gaussian tapered Airy disk model presented in \citealt{harper2018}, Airy in short) with chromatic side-lobes for the SKA-Mid and MeerKAT dishes. A more detailed discussion on the beam can be found in \autoref{sec:beam}.

The true level of \hi\ signal, obtained from a halo catalogue painted with a halo-\hi\ mass relation derived from a galaxy evolution semi-analytical model. For more details on \hi\ IM simulation in the context of the SKAO see \citet{Ronconi01.2026.SKA}. The participants used a total of nine different pipelines to clean the data-cubes, ranging from different blind (PCA, fastICA and GMCA) and non-blind (parametric fitting) source separation algorithms.
The results were analysed not only in terms of their absolute performance but also in terms of relative performance in order to understand better their weaknesses and strengths.

\begin{figure}
\centering
   \includegraphics[width=0.45\columnwidth]{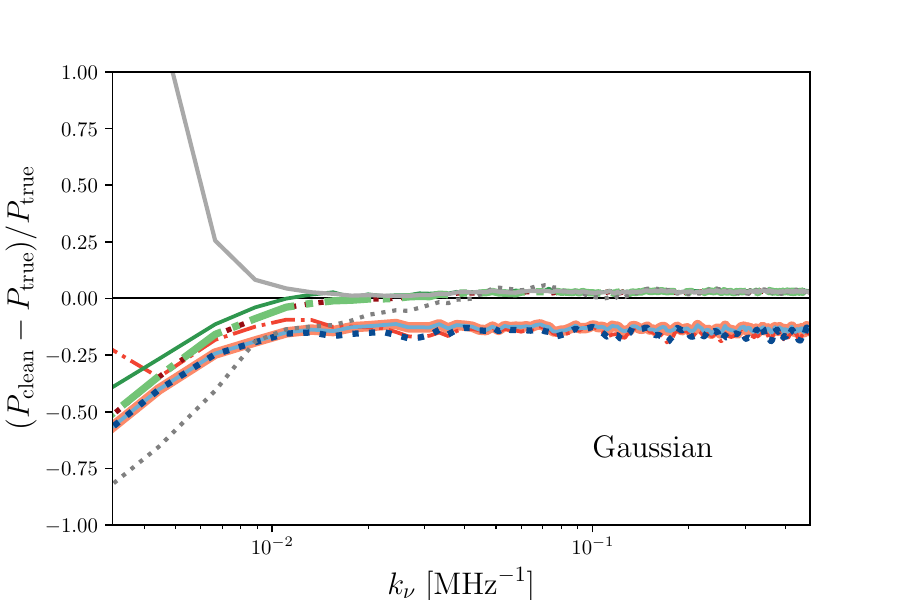}
\includegraphics[width=0.45\columnwidth]{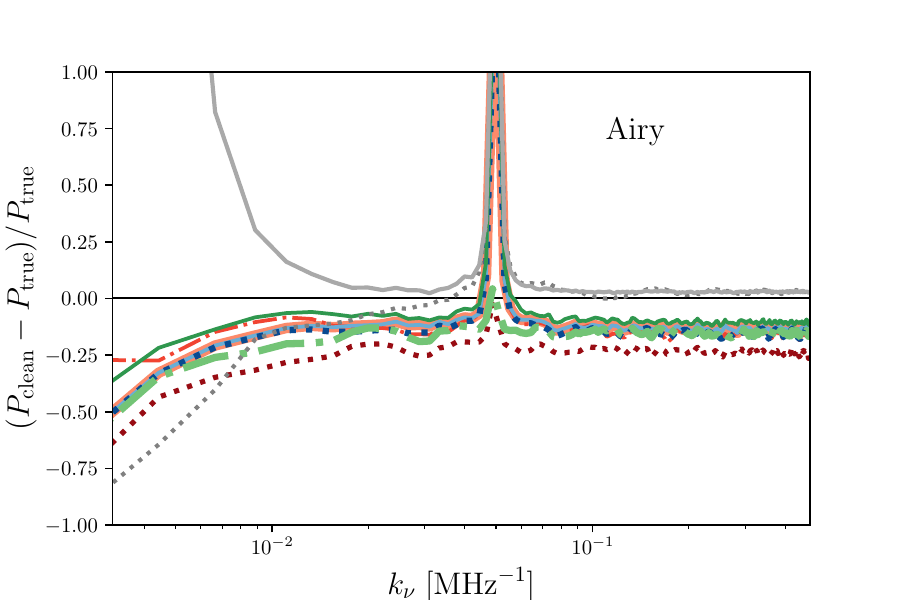}\\
    \includegraphics[width=0.5\columnwidth]{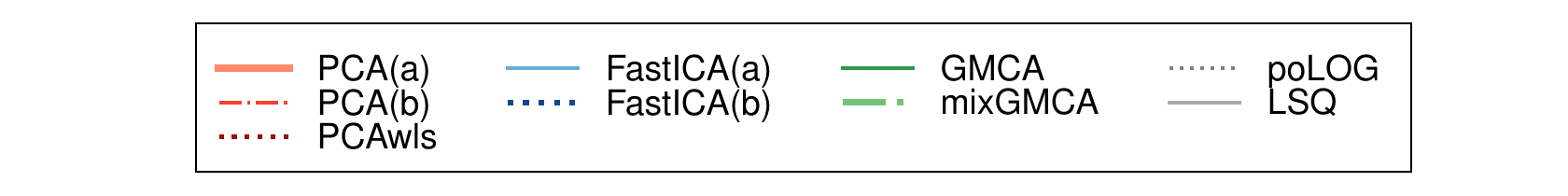}
   \caption{Comparison of the estimator $(P_{\rm clean}-P_{\rm true})/P_{\rm true}$, where $P_{\rm clean}$ is the power spectrum of the residual maps  for the various cleaning methods, while $P_{\rm true}$ is the input signal and noise. The left panel show the Gaussian beam case while the right panel show the result for a realistic beam model (Airy, see \citealt{harper2018})}
   \label{fig:unblind_res_pk}
\end{figure}

The results pointed out that, even among similar methods, subtleties related to each specific implementation can lead to substantial differences in the cleaning performance, and that the choice $N_{\rm fg}$ is not easily deducible and objective without extra prior information on the signal. In \autoref{fig:unblind_res_pk}, we report the performances of the various methods using the line-of-sight (los) power spectrum, a powerful statistic for residual structure in frequency. Blind methods are generally capable of recovering within 20\% the input power spectra in the frequency range (left panel) and spatial scales with the least beam suppression, for a simple Gaussian beam. The right panel shows instead the case of a Airy beam whose frequency dependence is fitted on MeerKAT's beam measurements. The presence of residual structure to to the coupling of the beam structure with spatial variation of the foreground was anticipated in \citet{Matshawule2021} and will be discuss in more detail in the next section. 

\begin{figure}
      \includegraphics[width=0.5\columnwidth]{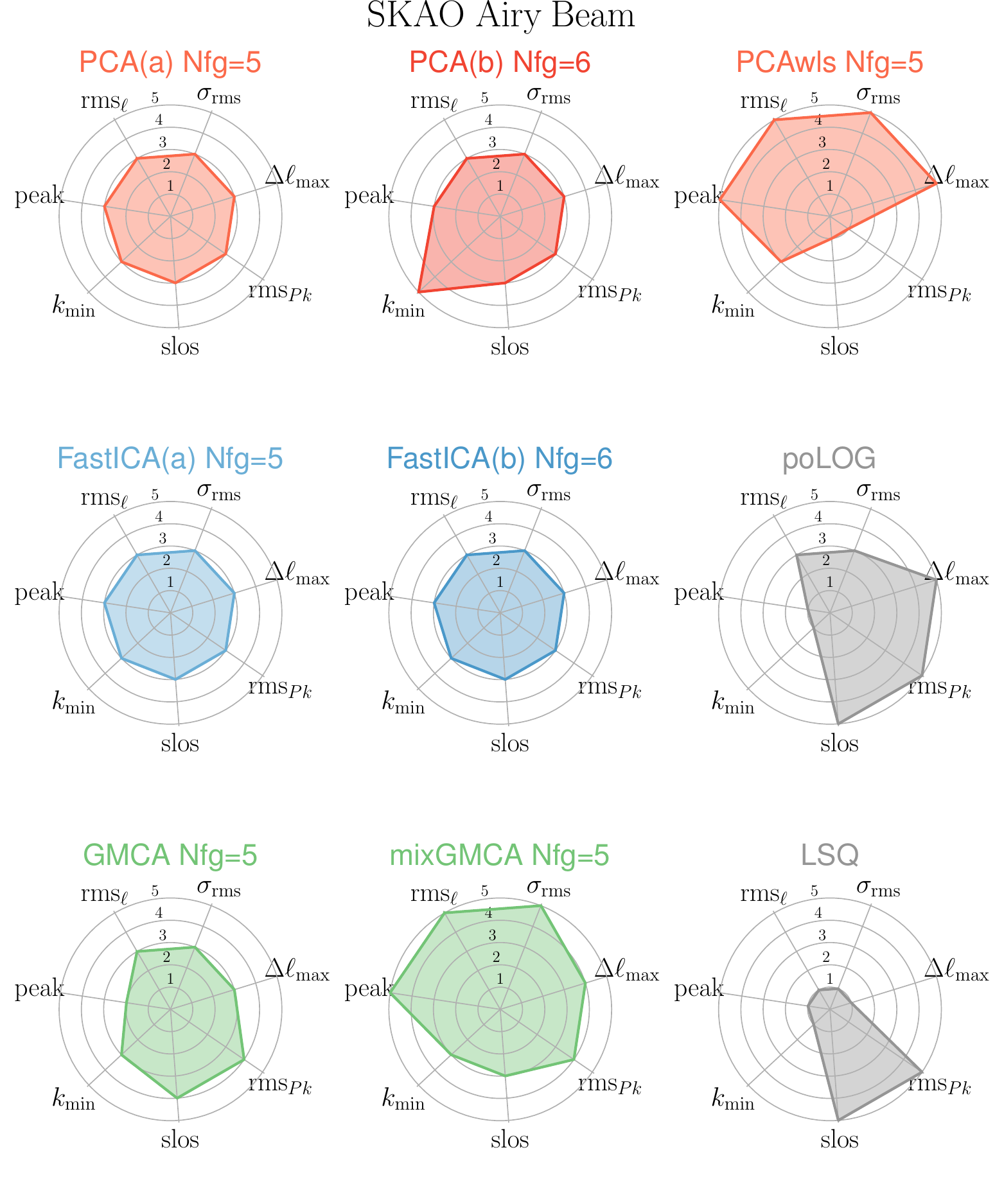}\hspace{0.5cm}
       \includegraphics[width=0.5\columnwidth]{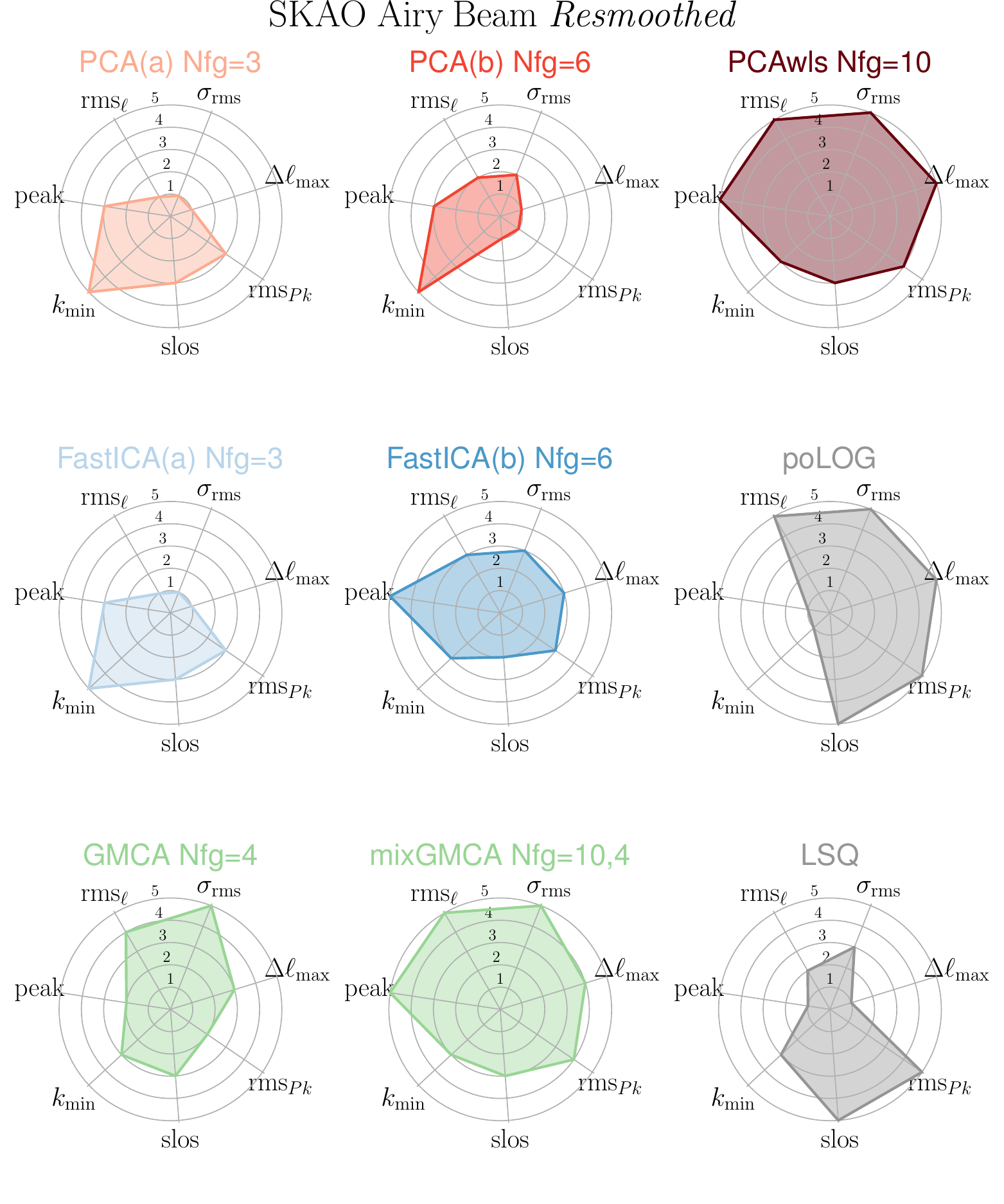}
   \caption{Radar charts showing the performance of the various cleaning methods participating in the data challenge presented in \citet{Spinelli2021} (larger covered areas correspond to better performing methods). Mock data are assuming a SKA-Mid \hi\ IM survey. See text for details.}
   \label{fig:spiderSKAO}
\end{figure}

Aiming at a detailed comparison among methods, the performances in reconstructing the true signal have been summarized in metrics defined for both the los power spectrum and the angular power spectra at each frequency/redshift (see \citealt{Spinelli2021} for details). These have been reported in the radar charts in \autoref{fig:spiderSKAO}: a better performing method corresponds to a more extended chart. Methods are colour-coded as in \autoref{fig:unblind_res_pk}. For each blind method, the number of subtracted components $N_{\rm fg}$ is also reported, and the intensity of the colour is scaled proportionally (darker colour corresponds to higher $N_{\rm fg}$). Note that mixGMCA is associated with two different $N_{\rm fg}$ the first for the largest scale PCA and the second for smaller scales GMCA. A similar multi-scale cleaning approach (see also \autoref{ss:pca}) has been recently successfully applied to real data in \citet{mPCA}.   

In the right panel the results are shown again for the realistic beam case but when the mock data are artificially resmoothed to a common low resolution via a Gaussian convolution. This is was a popular procedure for real data \citep[e.g.][]{Wolz:2015lwa} that was helping the cleaning procedure. We see here that the same is not true in simulations when there is a mismatch between the simulated telescope beam (Airy) and the one assumed for the resmoothing. A detailed discussion can be found also in \citet{Matshawule2021}.

\section{The effect of the telescope beam}\label{sec:beam}
\begin{figure}
\includegraphics[width=7.5cm]{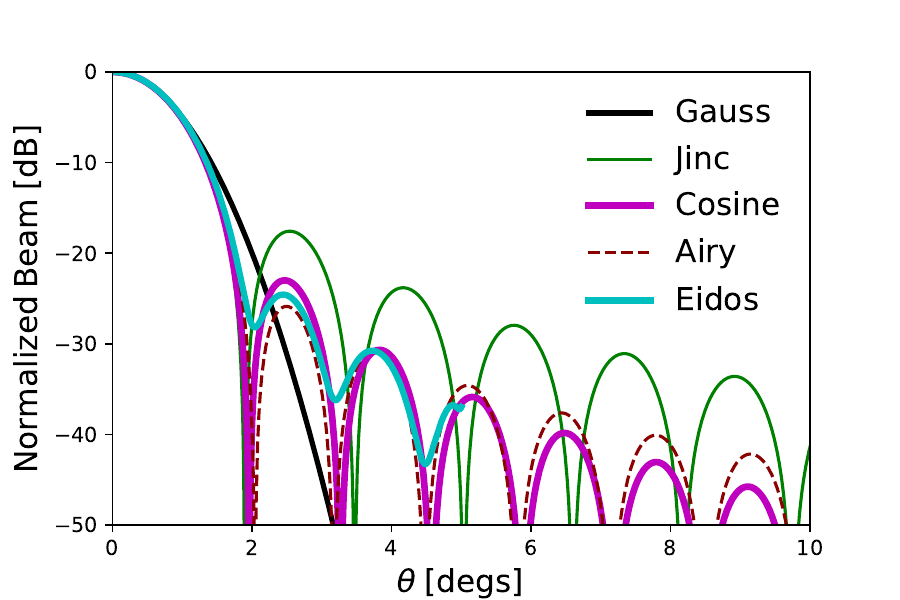}
\includegraphics[width=7.5cm]{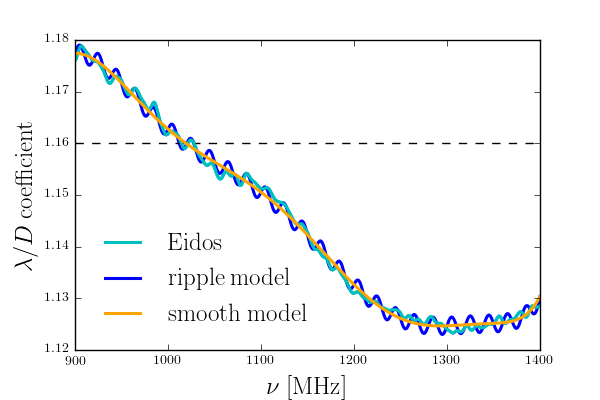}
\caption{\emph{Left}: Primary beam model at $950~\mathsf{MHz}$: Gauss model (black), Jinc model \citep{wilson2013} (solid green), the cosine model \citep{condonransom2016} (magenta), the gaussian tapered Airy disk used in \citet{harper2018} (dashed red), and the one obtained from the Eidos package presented in \citet{asad2019} (cyan).
\emph{Right}: azimuthally averaged FWHM of the MeerKAT/Eidos beam normalized by $\lambda/D$ as a function of frequency (solid cyan line) compared to the ripple model (solid blue line) which is composed of a sinusoidal oscillation on top of a smooth polynomial frequency dependence (solid orange line) - see equation~\ref{eq:fwhm_nu}. Figures taken from \citet{Matshawule2021}.}
\label{fig:beam_models}
\end{figure}

In single-dish IM, the telescope beam is often assumed to be Gaussian with a FWHM, $\Delta\theta$, given approximately by
\begin{equation}
 \Delta\theta \approx 1.16\frac{\lambda}{D},\ 
 \label{eq:fwhm}
\end{equation}
where $\lambda$ is the observed wavelength and $D$ the dish diameter. In reality, there are two main complications to this scenario: i) the beam has sidelobes that can be more or less pronounced (see the left panel of \autoref{fig:beam_models}) and can introduce structures in the maps coming from strong point sources picked up by that; ii) the beam can have a non-trivial chromaticity (see for example the structure in the right panel of \autoref{fig:beam_models}) that can severely affect component separation.

\subsection{Beam sidelobe structure}\label{ss:sidelobes}

Sidelobes can be problematic for IM experiments. For example, the extragalactic point sources which are located in the sidelobes of the primary beam pattern, can show up as oscillations in the measured multifrequency angular power spectrum (MAPS) $C_\ell (\Delta\nu)$ \citep{ghosh1}, which we will describe in more detail in \autoref{sec:MAPS}. These oscillations are extremely difficult to model and remove from the measured $C_\ell (\Delta\nu)$ but can be mitigated using window functions. This idea is implemented in the Tapered Gridded Estimator (TGE) again described in \autoref{sec:MAPS}. We refer the interested reader to \citet{ghosh2, samir14, samir16} for a detailed description.

In the specific context of blind component separation for single-dish IM, the results of \citet{Matshawule2021} seems to suggest that sidelobe structure of the primary beam do not cause major problems per se if there is no complex chromaticity in the beam. This can be seen in \autoref{fig:fgrm_ripple} (black and yellow lines), where, until the variation of the FWHM is smooth with frequency, the level of the point source contamination does not change the performance of the foreground cleaning.

\subsection{Primary beam chromaticity}\label{ss:ripple}

Measurements of the MeerKAT/Eidos \citep[e.g][]{asad2019} beam shows that the FWHM exhibits a low-level frequency-dependent {\it ripple}. This effect can be seen clearly in the right panel of \autoref{fig:beam_models}, where the Eidos FWHM is normalized by $\lambda/D$. This ripple is caused by the interaction between the primary and secondary reflector of MeerKAT \citep{villiers2013} and can be important in the foreground cleaning as it will add extra structure to the frequency spectra. 

\citet{Matshawule2021} proposed a simple model for the combination of the smooth structure and the ripple shown in the right panel of \autoref{fig:beam_models} and valid for the MeerKAT L-band: a high order polynomial with superimposed a sinusoidal oscillation with period $T$ and amplitude $A$, 
\begin{equation}
      \centering
     \Delta\theta_r = \frac{\lambda}{D}\left(\sum_{d=0}^{8}a_d \nu^{d} + A\sin\Bigg(\frac{2\pi\nu}{T}\Bigg)\right).
     \label{eq:fwhm_nu}
\end{equation}
%The values of the parameters are summarized in a table in the paper.

\begin{figure}[h!]
\includegraphics[width=7cm]{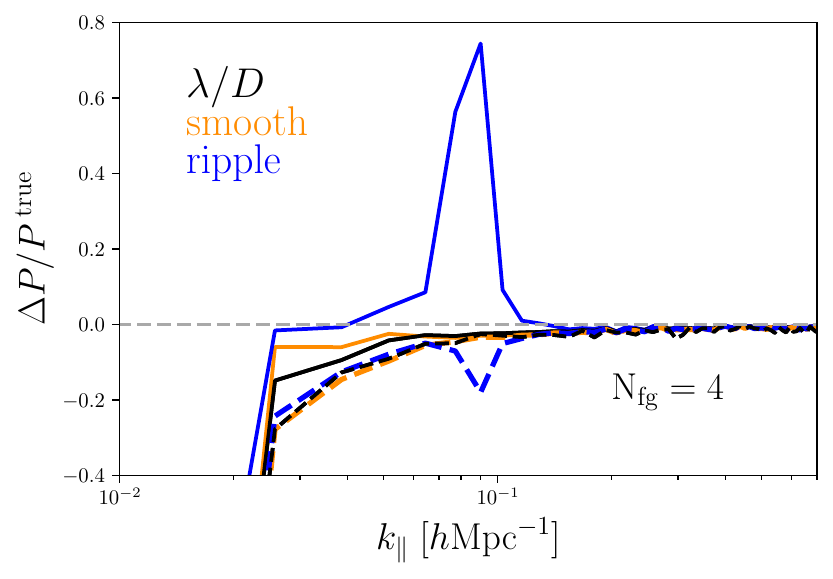}
\includegraphics[width=7.3cm]{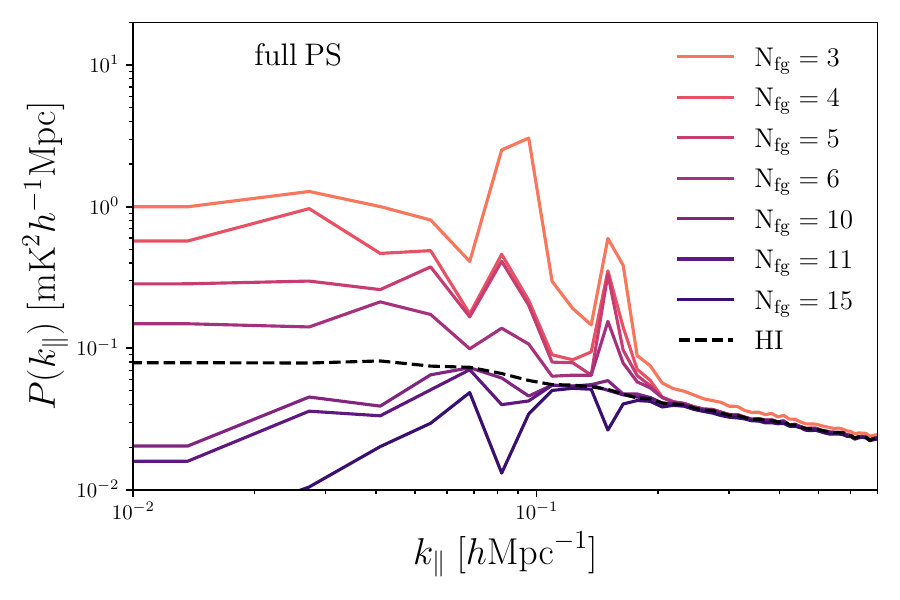}
\caption{\emph{Left.} Relative difference between the foreground cleaned (with $N_{\rm fg}=4$) radial power spectrum and the true input signal for the different FWHM models: $\frac{\lambda}{D}$ in black, the smooth model in orange and the ripple model in blue. Results are shown for the worst and best-case scenario for the point source contamination, considering the full catalogue (solid lines) or only point sources with a flux cut of $100$~mJy (dotted lines), respectively. Figure taken from \citet{Matshawule2021}. \emph{Right:} Difference, for a mock \hi\ observation with synchrotron and strong point sources, between the cleaned radial power spectrum and the input when also beam asymmetries are considered. Figure taken from \citet{MatshawulePhD}.}

 \label{fig:fgrm_ripple}
 \end{figure}

This type of bean chromaticity complicates component separation appearing as a residual oscillation in the ``cleaned" signal. The combined effects of sidelobes and chromaticity has been studied for PCA in \citet{Matshawule2021} in the context of simulations, and the effect can be distinctively seen in the structure of the line-of-sight power spectrum in the left panel of \autoref{fig:fgrm_ripple}. More recently, the effect has also been discussed for other cleaning techniques such as GNILC (see \autoref{ss:gnilc}, \citet{DeCaro-2025}) and Deep Learning (see \autoref{ss:ai}, \citet{nlg+22}).

Emission line stacking analysis of the MeerKLASS L-band deep-field survey confirmed the oscillation structure in observation for the first time \citep{2025ApJS..279...19C}. The chromatic systematics are found to be a convolutional effect on the data, and create correlations in the \hi\ signal which requires robust covariance estimation to model. Joint inferences on the systematics as well as the \hi\ signal can help constrain the effect of beam chromaticity.

Note that the ripple in the beam width is expected to be asymmetric between the vertical and the horizontal direction \citep{asad2019}.
The combination of beam asymmetries and sky rotation could result in a superposition of sine waves which will leak the ripple across more scales while possibly reducing its overall amplitude.
This can be seen in simulations when a mock \hi\ observation is convolved with Eidos beam maintaining the asymmetries and following a scanning strategy \citep{MatshawulePhD}. The right panel of \autoref{fig:fgrm_ripple} shows the more complex residual structure found especially in presence of strong point sources.

\subsection{Joint deconvolution and blind source separation on the sphere (SDecGMCA)}\label{ss:SDecGMCA}

As discussed in the previous section, a frequency-dependent beam can blur the observed sky maps in a way that complicates component separation and applying standard blind component separation methods to these cases can lead to spectral distortions or angular artifacts, ultimately biasing the recovery of the cosmological \hi\ signal. To mitigate these effects, an additional deconvolution step could be incorporated, giving rise to the joint deconvolution and blind source separation problem (DBSS). 

In this context, \autoref{eq:systosolve} can be extended to account for the convolution with the beam:
\begin{equation}
\mathbf{X}_{\nu} = (\mathbf{A}_{\nu} \mathbf{S}) \ast \mathbf{H}_{\nu} + \mathbf{N}_{\nu},
\label{eq:SDecGMCA_main_equation}
\end{equation}
where $\mathbf{X}_{\nu}$ denotes the observed maps at frequency $\nu$, $\mathbf{A}_{\nu}$ is the mixing matrix, $\mathbf{S}$ the matrix of source components, $\mathbf{H}_{\nu}$ the frequency-dependent beam operator, and $\ast$ denotes convolution.

The presence of a spatially smoothing and frequency-dependent operator—often ill-conditioned or even non-invertible—complicates the separation. The DBSS framework addresses this by jointly estimating both the sources and the mixing matrix while accounting for the beam effects. One such method is SDecGMCA, an extension of the GMCA approach (see \autoref{sec:gmca}), initially introduced by \citet{jiang2017} for sky patches and generalized to full-sky analysis in spherical harmonics by \citet{carloni2021}\footnote{\url{https://github.com/RCarloniGertosio/SDecGMCA}}.

SDecGMCA reformulates the problem in the spherical harmonic domain, where convolution becomes a simple multiplication:
\begin{equation}
    \hat{\mathbf{X}}_{\nu}^{\ell, m} = \hat{\mathbf{H}}_{\nu}^{\ell}\,\mathbf{A}_{\nu}\,\hat{\mathbf{S}}^{\ell, m} + \hat{\mathbf{N}}_{\nu}^{\ell, m},
\label{eq:SDecGMCA_spherical_harmonic_domain}
\end{equation}
with $(\ell, m)$ denoting the spherical harmonic coefficients, and where isotropy of the beam implies that $\hat{\mathbf{H}}_{\nu}^{\ell}$ depends only on $\ell$.

The algorithm minimizes a cost function that combines data fidelity with a sparsity-promoting prior on the sources:
\begin{equation}
    \underset{\mathbf{A}, \hat{\mathbf{S}}}{\rm min} \ || \mathbf{\Lambda} \odot ( \hat{\mathbf{S}}\,\mathcal{F}^{\dagger}\,\mathbf{\Phi}^{\sf T} ) ||_{1} \
    + \sum_{(\ell, m) \in \mathcal{D}} ||\hat{\mathbf{X}}^{\ell, m} - {\rm diag}(\hat{\mathbf{H}}^{\ell})\,\mathbf{A}\,\hat{\mathbf{S}}^{\ell,m} ||_{2}^2,
\label{eq:SDecGMCA_cost_function}
\end{equation}
where $\mathbf{\Lambda}$ contains thresholding parameters, $\odot$ denotes the element-wise product, $\mathbf{\Phi}$ is a wavelet transform, and $\mathcal{F}^{\dagger}$ denotes the inverse spherical harmonic transform (see \citet{Gkogkou2026} for more details).

The iterative algorithm alternates between: i) estimating $\mathbf{S}$ (sources) for fixed $\mathbf{A}$, using a regularized least-squares solution that enforces sparsity and stabilizes the inversion of the beam; ii) updating $\mathbf{A}$ for fixed $\mathbf{S}$ via least-squares; iii) refining the regularization parameters across iterations to ensure convergence and enhanced separation quality.

To evaluate the applicability of SDecGMCA to recovering the \hi\ cosmological signal, the method was applied to a full-sky simulated data set in \citet{Gkogkou2026}. The simulations include cosmological \hi\ emission alongside dominant astrophysical foregrounds—such as synchrotron radiation, free-free emission, and unresolved point sources. It also incorporates realistic instrumental noise based on MeerKAT-like specifications. In order to assess the robustness of the method in the presence of frequency-dependent instrumental distortions, three beam models of increasing complexity were tested: 
a Gaussian-degraded beam with constant resolution across frequencies, a Gaussian-evolving beam with a smooth, frequency-dependent width (often assumed in the studies presented in this chapter), and the Gaussian-oscillating beam with the frequency structure studied in \citet{Matshawule2021}.

\begin{figure}[h!]
    \centering
    \begin{minipage}{0.48\textwidth}
        \centering
        \includegraphics[width=\textwidth]{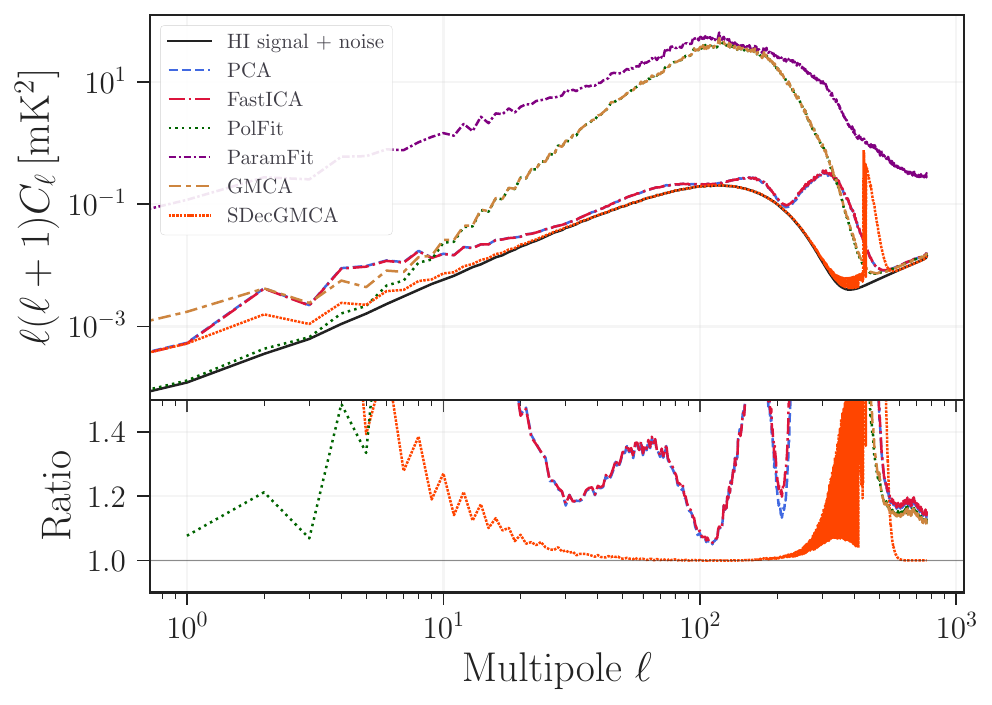}
    \end{minipage}
    \hfill
    \begin{minipage}{0.48\textwidth}
        \centering
        \includegraphics[width=\textwidth]{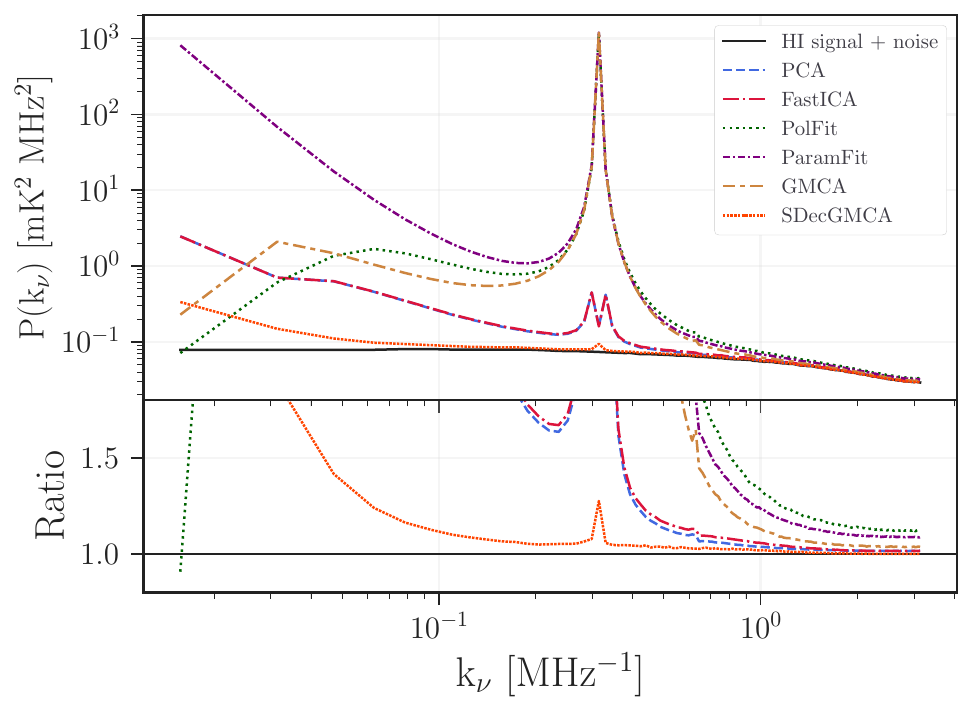}
    \end{minipage}
    \caption{Angular (left) and frequency (right) power spectrum of the reconstructed \hi\ IM signal using different methods, each represented by a distinct colour. The bottom panels display the ratio of the reconstructed to the true \hi\ power spectrum for each method.}
    \label{fig:results_oscillating_beam}
\end{figure}

The results (see \autoref{fig:results_oscillating_beam}) indicate that SDecGMCA consistently outperforms both standard blind subtraction techniques and parametric fitting methods in the presence of complex, frequency-dependent beam distortions, particularly under the oscillating beam model. Improved recovery is achieved for both the frequency and angular power spectra of the \hi\ cosmological signal. Furthermore, several reconstruction artifacts previously reported in \citep{Spinelli2021}, including the systematic offset and the spurious peak in the frequency power spectrum, are substantially reduced or entirely suppressed.
This result demonstrates the potential of the joint beam deconvolution and cleaning although some improvement are needed to make the algorithm mature for the data. For example, the method has been tested on full-sky simulated observations and a next step would be to incorporate a masking strategy within the framework. 

\section{Other approaches to component separation}\label{sec:news}
\subsection{Foreground removal via the multi-frequency angular power spectrum}\label{sec:MAPS}

The idea is based on the distinctly different spectral behavior of the 21 cm signal as compared to the foregrounds. The signal is expected to be predominantly localized within a typical $\Delta \nu $ range of  $ 0 - 0.5 \, \rm{MHz}$, and its amplitude drops substantially and is close to zero at large $\Delta \nu$ \citep{Bharadwaj2001a, Bharadwaj2001b, Bharadwaj2005}. On the other hand, the foregrounds generally vary smoothly and remain correlated over a large $\Delta \nu$ range. Following \citet{ghosh1}, it is possible to adopt a two-step procedure to remove the foregrounds from the estimated $C_\ell (\Delta\nu)$. First, one can identify  a characteristic length scale $\left[ \Delta \nu \right]$ such that in the range   $\Delta \nu > \left[ \Delta \nu \right]$ the  21 cm signal is predicted to be  negligible, and model $C_\ell (\Delta\nu)$ in this range  as a combination of foregrounds $\left[C_{\ell}(\Delta\nu)\right]_{\rm FG}$ and noise
\begin{equation}
    C_{\ell}(\Delta\nu) = \left[C_{\ell}(\Delta\nu)\right]_{\rm FG} + [\textrm{Noise}] \,.
    \label{eq:polyfitmodel} 
\end{equation}
The procedure is then to use the $C_\ell (\Delta\nu)$  measured in the range $\Delta \nu > \left[ \Delta \nu \right]$ to estimate  the foreground model $\left[C_{\ell}(\Delta\nu)\right]_{\rm FG}$. 
  
In the second step, the foreground model can be extrapolated to predict $ \left[C_{\ell}(\Delta\nu)\right]_{\rm FG}$ in the range $\Delta \nu \leq \left[ \Delta \nu \right]$ and subtract this out from the measured $C_\ell (\Delta\nu)$. 
Restricting the subsequent analysis to $\Delta \nu \leq \left[ \Delta \nu \right]$, we expect the residual 
\begin{equation}
    \left[C_{\ell}(\Delta\nu)\right]_{\rm res} = C_{\ell}(\Delta\nu) - \left[C_{\ell}(\Delta\nu)\right]_{\rm FG} 
    \label{eq:residual}
\end{equation}
to be a combination of the 21 cm signal and noise
\begin{equation}
     \left[C_{\ell}(\Delta\nu)\right]_{\rm res}= \left[C_{\ell}(\Delta\nu)\right]_{\rm T} + [\textrm{Noise}]
\end{equation}
and we use this to constrain the  21 cm signal $\left[C_{\ell}(\Delta\nu)\right]_{\rm T}$.

\begin{figure}
\centering
\includegraphics[width=11cm,height=11cm,keepaspectratio]{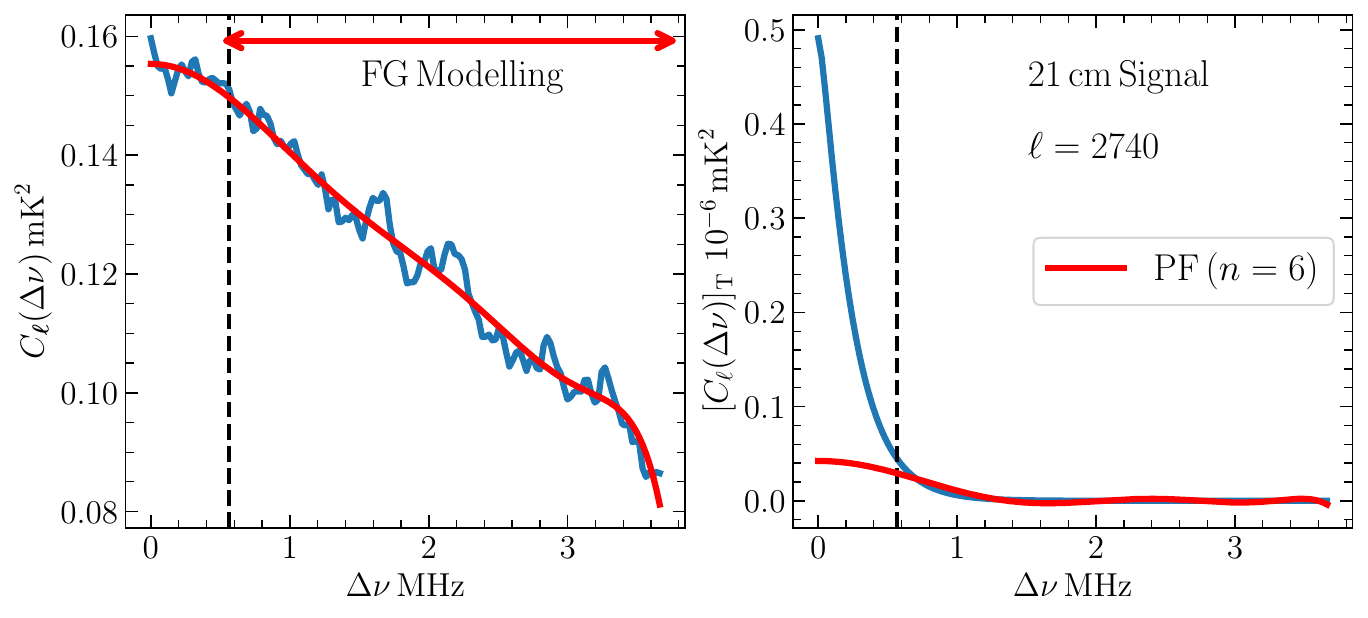}
\caption{The left and the right panels respectively show (blue solid line) the measured $C_{\ell}(\Delta\nu)$ and $[C_{\ell}(\Delta\nu)]_T$ for a representative $\ell = 2740$. The black dashed vertical lines show the chosen characteristic length scale $[\Delta\nu]_{0.1}$, where the amplitude of $[C_{\ell}(\Delta\nu)]_T$ falls to $10\%$ of the maximum correlation, and therefore there is a negligible amount of the signal in the range $\Delta\nu > [\Delta\nu]_{0.1}$ . The red solid curve shows the optimally found  $\left[C_{\ell}(\Delta\nu)\right]_{{\rm FG}}$ which minimizes the residual $ [C_{\ell}(\Delta\nu)]_{\rm{res}}$ in the range $\Delta\nu < [\Delta\nu]_{0.1}$. The dashed curves and the red solid curve in the right panel shows $[C_{\ell}(\Delta\nu)]_{TP}$ the best-fit polynomials for $[C_{\ell}(\Delta\nu)]_T$ for the same values of $n$.}
\label{fig:fg_removal_maps}
\end{figure}

In case of actual data (such as \citealt{AE23b, Elahi2024}), it has been found that the $\Delta\nu$ dependence exhibits a smooth, slowly varying pattern, superimposed on which we have a combination of rapid and slow oscillations. The aim of this foreground removal approach is to use the range $\Delta \nu > \left[ \Delta \nu \right]$ to model the smooth, slowly varying $\Delta\nu$ dependence. It turns out that modeling the oscillatory components either leads to extrapolation biases or a very large signal loss \citep{Elahi2024}. 
The foreground removal technique is implemented in two different approaches. In the first approach, polynomial fitting (PF), one can model the foregrounds using an even-order polynomial. 
We use maximum likelihood to estimate the best-fit polynomial coefficients and their error covariance,  and use these to obtain the best-fit foreground and the foreground modeling errors in the range $\Delta \nu \leq \left[ \Delta \nu \right]$. 
Note that we have restricted the order of the polynomial to model only the smooth, slowly varying component from the measured  $C_{\ell}(\Delta\nu)$. 
In the second approach, one can use Gaussian Process Regression (GPR; \citealt{RW}), which is non-parametric, to model and predict the foreground. This was first implemented in \citet{AE23b}, where the reader can find more details. In GPR, we model $\left[C_{\ell}(\Delta\nu)\right]_{\rm FG}$ as a Gaussian Process (GP)
\begin{equation}
    \left[C_{\ell}(\Delta\nu)\right]_{\rm FG} \sim \mathcal{GP} \left[ 0,\, k_{{\rm fg}}(\Delta\nu_m, \Delta\nu_n) \right] \,,
    \label{eq:FGmodel_GP}
\end{equation}
with a zero mean  and  covariance function (kernel)  $ k_{{\rm fg}}(\Delta\nu_m, \Delta\nu_n)$. The idea is to use the $C_{\ell}(\Delta\nu)$  measured in the range $\Delta \nu > \left[ \Delta \nu \right]$ to estimate the kernel, which models the smooth, slowly varying foreground component $ \left[C_{\ell}(\Delta\nu)\right]_{\rm FG}$. We subsequently use this kernel to predict $ \left[C_{\ell}(\Delta\nu)\right]_{\rm FG}$ in the range $\Delta \nu \leq  \left[ \Delta \nu \right]$. We have tried  different possible forms for the kernel, and found that the polynomial  kernel 
\begin{equation}
    k_{{\rm fg}}(\Delta\nu_m, \Delta\nu_n) = c_1 \, (\Delta\nu_m \cdot \Delta\nu_n + b )^{P}  
    \label{eq:kernel}
\end{equation}
is well suited for our analysis. Here, the constants $c_1$ and $b$ are hyper-parameters  whose  optimal value  have been determined  by maximizing the log marginal likelihood of the GP, and  $P$, which is not a hyper-parameter, denotes the order of the polynomial kernel.  The polynomial kernel is found to model the overall smooth $\Delta \nu$ dependence expected of the foregrounds and causes a minimal signal loss. We have considered different values of $P$ and found that larger values provide a better fit to the foregrounds in the range $\Delta \nu > [\Delta \nu]$ but at the cost of a larger signal loss. Considering this, we have used $P = 2$ or $3$ which provides reasonable choice for \hi\ IM analysis. 
The GPR predicts the foregrounds $\left[C_{\ell}(\Delta\nu)\right]_{\rm FG}$ and also the foreground modeling errors in the range $\Delta \nu \leq [\Delta \nu]$. For both PF and GPR, we add the variance of foreground modeling errors to the estimated system noise variance to estimate the total error variance for $\left[C_{\ell}(\Delta\nu)\right]_{\rm res}$.

\subsection{High dimensional Bayesian modeling}
Another alternative to blind foreground removal is to attempt to model the foregrounds and instrumental response directly, in order to subtract the non-cosmological components. The high dynamic range between the foregrounds and 21 cm signal makes this challenging, as does the general complexity of instrumental models, and incompleteness of sky models. Any inaccuracy in the sky or instrument model will result in a model error, with even sub-percent errors able to swamp the cosmological signal. Bayesian statistical methods, which not only estimate the best-fit model parameters from the data, but also describe their statistical distribution, can be used to handle the model uncertainty. The uncertainty on the the foreground and systematics can then be propagated into the subsequent analysis steps, mitigating difficulties such as signal loss and other forms of biased results that are present in blind methods. The sky and instrument models need to be flexible enough to incorporate the true unknown behavior of these components. For example: varying source fluxes and power-law spectral indices for the point sources; diffuse foreground emission modeled per-pixel, or in Fourier or spherical harmonic bases. The spectral behavior of the latter can be specified using basis functions such as the principal components found by blind foreground methods, or physics-based approaches such as explicit modeling or moments.

The ever-increasing quantities of data expected for the SKAO necessitates these more sophisticated statistical methods to remain computationally tractable. With Bayesian sampling, models with as few as 100 parameters can begin to suffer from the \textit{curse of dimensionality}: a severe reduction in sampling speed when using more conventional algorithms such as Markov Chain Monte Carlo (MCMC). These performance limitations can be mitigated but not solved using  more sophisticated strategies as Hamiltonian Monte Carlo \citep[HMC]{Duane_1987,Betancourt_2017}.

One approach which can be resilient to highly-dimensionality is Gibbs sampling, which has been used to great effect in other fields, for example CMB experiments in \citet{Eriksen_2008a} which applies the work of \citet{Jewell_2004} and \citet{Wandelt_2004} to simulated WMAP data, and to observational data in \citet{Eriksen_2008b}. Further examples include \citet{beyondplanck_I_2023}, which applies Gibbs sampling to observations from the Planck Low Frequency Instrument (LFI); the reprocessing of WMAP data through an extension of the BeyondPlanck framework in \citet{2023A&A...679A.143W}; and \citet{2024arXiv240810952W}, which presents the first application of this framework to COBE Diffuse Infrared Background Experiment (DIRBE) data in Cosmoglobe DR2. This method allows for the constraint of millions of free model parameters on tractable timescales, assuming some careful modeling choices are made. 
%For example, in the DR1 analysis of \citet{2023A&A...679A.143W}, drawing a single Gibbs sample for the WMAP+LFI data took 812 CPU hours. On modern high-performance computers, this translates to approximately one month needed to draw a statistically significant number of samples ($>1000$), although alternative sampling methods would generally scale far less favourably for such a high-dimensional joint posterior.
Rather than sampling from the model's full posterior space, Gibbs sampling instead samples from subsets of this space. By choosing linear models, and ensuring Gaussian priors, these conditional distributions \citep{Lynch2007} (or {\it Gaussian Constrained Realisations}, GCR) can be sampled from iteratively, which is more efficient than sampling from the full posterior. Furthermore, as result of the model's ensured linearity, solutions can be found through the direct solutions of linear equations, with no need to evaluate likelihood functions. 
%With the advent of radio interferometers containing on the order of 10 to 100 dishes (and 1000 for SKA-Low, in future) which output measurements into millions of pixels at high cadences, Gibbs sampling and GCR allows for modelling of the 21 cm signal with associated statistical uncertainties, but with performance which remains tractable despite the immense data volumes present in modern-day cosmological experiments.

There are numerous possible applications for Bayesian methods such as this to future SKA interferometric observations for the epoch of reionization \citep[e.g][]{Glasscock_2024,Kennedy_2023}. An application which specifically addresses single-dish IM can be found in \citep{Murphy_2026} where the authors achieve a jointly model the 21 cm signal and foregrounds on a timescale of $<1\:\rm hour$, despite the model containing over 3 million free parameters.

\subsection{Deep Learning}\label{ss:ai}
In recent years, machine learning techniques have been proposed by many studies as a novel approach for foreground removal. To some degree, subtracting foreground residuals from \hi\ maps is similar to image denoising in the deep learning field. For example, \citet{mlv+21} applied a deep convolutional neural network (CNN) with a U-Net architecture to simulated single dish data. Compared with PCA, they showed that U-Net can effectively subtract foregrounds from the signals with improved accuracy at intermediate radial scale ($0.025<k_{||}<0.075$ h\,Mpc$^{-1}$). \citet{nlg+22} tested U-Net on simulated MeerKLASS data and found that the U-Net can effectively clean the foreground with both Gaussian and Cosine beam effects. Following \citet{nlg+22}, \citet{gln+22} demonstrated that the U-Net is able to eliminate  polarization leakage effects while subtracting 21 cm signal from foregrounds.

While deep learning shows promise for foreground removal in simulations, caution is warranted when applying these methods to real data due to limitations in reliability and generalization. The effectiveness of deep learning models is highly dependent on the training data, which is predominantly based on simulations. However, discrepancies between simulated and real data, stemming from differences in sky model assumptions or instrumental systematics, can compromise the accuracy of foreground subtraction. For example, \citet{cbe+24} evaluated the performance of a  U-Net using various sky simulation models for SKA-Mid Band\,1.  They found that the U-Net performs comparably to PCA only when the training and the test dataset originate consistently from the same sky model. When different models were used for training and testing, the network failed to accurately subtract foregrounds.  Moreover, their study demonstrated that the U-Net's ability to handle systematics such as  frequency-dependent Gaussian beams and bandpass fluctuation, depends heavily on the inclusion of these systematics in the training phase.  \citet{cbe+24} cautions that applying U-Net to real data requires care in two respects: simulated training data may not fully capture the complexity of observations, and the network relies on prior knowledge of systematics being included during training for accurate performance.

Nevertheless, deep learning offers an independent approach to 21 cm foreground removal which, when applied appropriately, can yield complementary results that serve as a valuable cross-check against traditional methods. Currently, deep learning techniques for 21 cm foreground removal  often rely on a  pre-processing step using  traditional methods  to reduce the large dynamic range between foregrounds and the signal prior to refined signal reconstruction. Future developments may focus on training the network to directly handle the foreground-to-signal ratio, potentially eliminating the need for pre-processing.  Additionally, integrating physical priors, such as the expected power spectrum of the 21 cm signal or foregrounds, into the network architecture might enhance the model's robustness to variations in real data.

\section{Signal loss}\label{sec:loss}

Whilst there is a large effort to limit foreground residuals entering $\boldsymbol{\mathsf{N}}$ from \autoref{eq:systosolve}, causing \textit{additive} contamination, an equally important concern is \hi\ signal leakage into the removed $\boldsymbol{\mathsf{A}}\boldsymbol{\mathsf{S}}$ term, which causes a suppression in the measured \hi\ power spectrum, known as \textit{signal loss}.

\subsection{Transfer function}\label{ss:TF}

Constructing a foreground transfer function $\mathcal{T}(\boldsymbol{k})$ is seen as a promising method for correcting signal loss. This quantifies how the true \hi\ power spectrum is modulated by the cleaning algorithm. It is defined such that
$P_{\rm clean}(\boldsymbol{k})\,{=}\,\mathcal{T}(\boldsymbol{k})\,P_{\rm true}(\boldsymbol{k})$, and if an effective means can be devised to provide a transfer function estimate, $\widehat{\mathcal{T}}(\boldsymbol{k})$, then the signal can be corrected for at the power spectrum level, i.e. this is not a direct reconstruction of the map's attenuation. The unbiased power spectrum used for cosmological inference is thus simply $P_{\rm rec}(\boldsymbol{k}) = P_{\rm clean}(\boldsymbol{k}) / \widehat{\mathcal{T}}(\boldsymbol{k})$.

In practice, $\widehat{\mathcal{T}}(\boldsymbol{k})$ is obtained empirically by injecting mock maps of \hi\ into the real contaminated data, repeating the same blind foreground cleaning, and measuring the fractional suppression of the recovered mock power \citep{Switzer:2015ria, Cunnington:2023jpq}. Using the true data ensures that instrumental effects, noise correlations, and realistic foreground complexity are faithfully represented. The ratio of cleaned to input mock power spectra provides the estimate of the transfer function i.e.
\begin{equation}
    \widehat{\mathcal{T}}(k) = \Big\langle
        \frac{\mathcal{P}(\textbf{\textsf{M}}_{\rm clean}, \textbf{\textsf{M}})}{\mathcal{P}(\textbf{\textsf{M}}, \textbf{\textsf{M}})}
        \Big\rangle_{N_{\rm mock}} \,,
\end{equation}
where $\textbf{\textsf{M}}_{\rm clean}$ is the cleaned mock and real data combination, and $\textbf{\textsf{M}}$ is the original injected mock, uncleanded. This is averaged over multiple mocks ($N_{\rm mock}$) to yield a smooth, robust estimate of $\widehat{\mathcal{T}}(\boldsymbol{k})$ that captures both the scale dependence and anisotropy of the signal loss.

Tests with simulated observations in \citet{Cunnington:2023jpq} demonstrated the accuracy of this approach: even when a large number of foreground modes are removed, the reconstructed power spectra remain unbiased at the sub-percent level across the scales well sampled by the survey. These results validate the empirical transfer function correction as a practical and reliable procedure for recovering the true \hi\ clustering signal from aggressively cleaned data.

Further work in \citet{2025MNRAS.542L...1C}, found that the transfer function correction of signal loss is effectively a normalization of the power spectrum window function in the quadratic estimator formalism (see e.g. \citealt{2021MNRAS.501.1463K}). Through the window function, the effects of foreground cleaning can be understood as a signal attenuation as well as \textit{mode-mixing}. Robust inference using the \hi\ intensity mapping in the future thus requires accurate modelling of both effects in the power spectrum estimation as well as in the covariance.

\subsection{Parametrizing signal loss}\label{ss:parTF}
As an alternative to correcting for signal loss at the data level through a transfer function, it is possible to incorporate a phenomenological treatment into the model for the comparison with the measurements and introduce one (or more) nuisance parameters in the fit to constrain the impact of signal loss. Based on simulations, the transfer functions can be well described by a power law behavior: the most straightforward choice is therefore to include a kernel $\mathcal{D}_\mathrm{sl}$ with such a functional form within the model. In the most general expression, the kernel reads
\begin{equation}
    \mathcal{D}_\mathrm{sl}(k) = \left(\frac{k}{k_\mathrm{pivot}}\right)^\beta \;
\end{equation}
where $k_\mathrm{pivot}$ acts as a normalization point and  $\beta$, that can be left as free parameter, controls the scaling and strength of the effect. More complicated expressions, as broken power laws or kernels that separate the radial and transverse directions, can be applied as well. When including this correction, a generic model of the \textsc{Hi} power spectrum $P_{\textsc{Hi}}^\mathrm{th}(k)$ is modified as $P_{\textsc{Hi}}^\mathrm{th}(k) \, \to \, \mathcal{D}_\mathrm{sl}(k)P_{\textsc{Hi}}^\mathrm{th}(k)$. This parameterisation is capable of capturing possible contributions from residual foregrounds, although these are expected to be more under control: a positive value of $\beta$ corresponds to signal loss on large scales, with the severity increasing with $\beta$, whereas a negative value indicates an excess of signal. The main limitations of this approach are the possible degeneracies with other parameters of interest \citep[see e.g.][]{Cunnington:2020wdu,Soares:2020zaq} and the reduced constraining power resulting from the enlargement of the parameter space. On the other hand, it provides a viable alternative to reconstructing the signal itself and it enables a direct estimation of the impact of signal loss from the fits alone, without the need for simulations. This makes it particularly well suited for analyses characterized by a relatively small set of free parameters.

\section{Radio Frequency Interference}\label{sec:RFI}
Radio frequency interference (RFI) is emitted from various terrestrial and orbital sources, such as Global System for Mobile Communication (GSM) antenna tower or satellites and it is one of the biggest issue with an high dynamic range experiment like \hi IM.
Single-dish observation are particularly sensitive to RFI  due to their wide primary beam and sensitivity.  Even when observing far way from the main RFI frequency, secondary harmonics may still leak into the data \citep{harper2018}. Moreover, low level broad-band RFI is particularly difficult to remove. 
As the RFI environment continues to become more challenging, it is crucial in radio observation to find novel algorithms to recognise and possibly mitigate this contamination.

\citet{Engelbrecht2025} simulated RFI from the Radio Navigational Satellite System in MeerKAT L-band in terms of positioning and power seen through the telescope beam, and calibrated the simulation using data. Thanks to the calibration step, they extrapolated the possible contamination out of band, especially for the lower frequencies of interest for cosmology, finding that it seems possible to remove the potential bias with standard component separation methods. 

Even when recognised and flagged, RFI pose a severe challenge as it will cause missing data. It is fundamental to test the robustness of the cleaning step and the summary statistic estimation to this problem. \citet{Carucci:2020enz} have found that cleaning methods can deal with missing bands and the power spectrum estimation procedure described in \citet{Cunnington:2023jpq} seems also to be robust to missing data. 
Another way to deal with RFI flags is also to use the multi-frequency angular power spectrum described in \autoref{sec:MAPS} which easily deals with the missing frequency channels in the visibility data, which are flagged due to RFI. \citep{Bh18, Pal20}. One can indeed estimate the MAPS $C_\ell (\Delta\nu)$ first, and then Fourier transform it to estimate the power spectrum $P(k_{\perp}, k_{\parallel})$. In this approach, one can use only the available frequency channels, and it is not essential to compensate for the missing frequency channels.

\section{$1/f$ noise and its inclusion in map-making}\label{sec:pink}
$1/f$, pink or flicker noise is a systematic noise bias most notable within single dish data. It is caused by receiver gain fluctuations which, instead of being random are in fact correlated over time and result in a correlated multiplicative effect on the time-ordered-data which can sometimes be seen in the map domain as streaks which follow the path of the scanning strategy. For this reason it is also called stochastic gain or correlated noise. In \autoref{ss:pink}, we discuss the modelling of $1/f$ and its effect of cleaning in the context of single-dish observations and map-making pipelines.

The interplay between smooth and stochastic gain fluctuations, radiometer noise, and sky signal estimation forms, however, a tightly coupled inference problem in autocorrelation measurements. Treating these components in isolation, as discussed in \autoref{ss:pink}, can obscure important degeneracies and leads to systematic effects in high dynamic range observations. A recently proposed solution to this is presented in \autoref{sec: 1/f and map making}.

\subsection{The $1/f$ in IM single-dish observations}
\label{ss:pink}
 For measurements taken at a single frequency $1/f$ noise is parametrized as follows: 
\begin{equation}
S(f) = \sigma_n^{2} \left( 1 + \left(\frac{f_{k}}{f}\right)^{\alpha} \right),
\end{equation}
where $\sigma_n$ is the Gaussian instrumental noise, $f$ is the Fourier pair of the time elapsed across the time-ordered-data and the knee frequency ($f_{k}$) marks the frequency where the $1/f$ slope, which has a gradient of $\alpha$, intersects with the thermal noise level \citep{bsazy}. For a single experiment which measures time-ordered-data across multiple frequency channels, noise can be correlated across both time and frequency and so \citet{Harper181f} considered the following extension to the single channel $1/f$ noise:
\begin{equation}
{\bf{S}}(f, \tau) = \sigma_{n}^{2} \left( 1 + \frac{1}{K\delta \nu} \left(\frac{f_{k}}{f}\right)^{\alpha} \left(\frac{\tau_{0}}{\tau}\right)^{\frac{1-\beta}{\beta}} \right), \text{ where }
K = \int d \tau \, {\rm{sinc}}^{2}(\pi \delta \nu \tau) \left(\frac{\tau_{0}}{\tau} \right)^{(1-\beta)/\beta},
\end{equation}
%\begin{equation}
%{\bf{S}}(f, \tau) = \sigma_{n}^{2} \left( 1 + \frac{1}{K\delta \nu} \left(\frac{f_{k}}{f}\right)^{\alpha} \left(\frac{\tau_{0}}{\tau}\right)^{\frac{1-\beta}{\beta}} \right),
%\end{equation}
%where
%\begin{equation}
%K = \int d \tau \, {\rm{sinc}}^{2}(\pi \delta \nu \tau) \left(\frac{\tau_{0}}{\tau} \right)^{(1-\beta)/\beta},
%\end{equation}
with $\tau_{0} = (N_{\nu} \delta\nu)^{-1}$ and $N_{\nu}$ representing the total number of frequency channels. For this 2D parameterization it is the $\beta$ term which quantifies the frequency correlation; $\beta$ would be equal to 1 for an experiment where there were no correlations at all between the $1/f$ noise in each frequency channel, while the closer $\beta$ tends to zero the stronger the frequency correlations. Mitigating the $1/f$ noise bias is an important data reduction task for single-dish experiments and has been considered in the context of \hi\ intensity mapping for the SKA-Mid \citep{Chen20201f}, FAST \citep{hu21} and BINGO \citep{bingo_f}.

\citet{li21} placed estimates on the level of $1/f$ noise within the MeerKLASS survey data through dedicated MeerKAT observations of the South Celestial Pole (SCP). They found that the contamination from $1/f$ noise, which on average had knee frequencies of around 0.1\,Hz, could be reduced using a 2-mode PCA clean of the time-ordered-data down to the level of $f_{k} = 0.003\,$MHz. 

 \begin{figure}
\centering
{\includegraphics[width=0.6\linewidth]{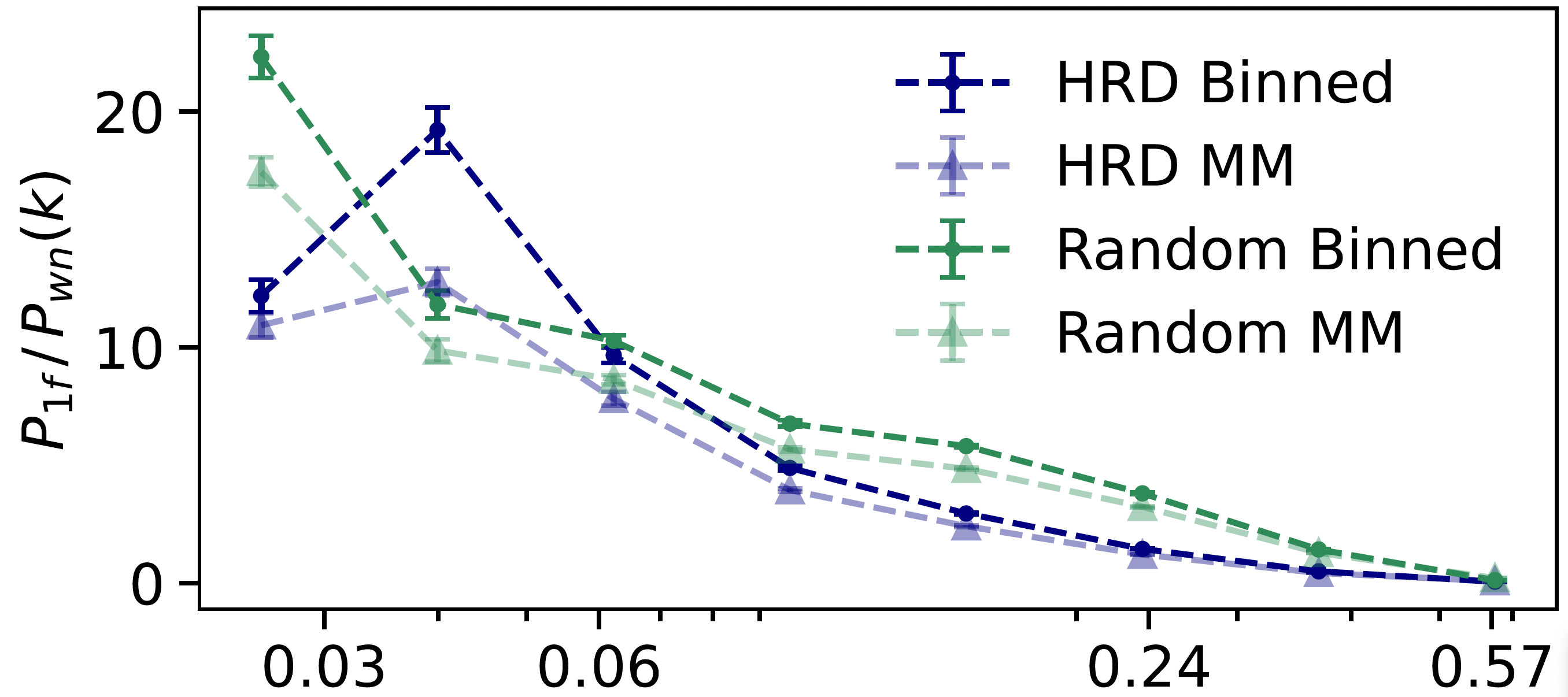}}\\
\caption{Spherically averaged power spectra for simulated $1/f$ noise above the instrumental noise level. Two different scan strategies are considered: HRD and random. Forming sky maps from the time-ordered data was done with (MM) and without (Binned) the inclusion of the noise covariance matrix. Figure adapted from \citet{irfan24}.}
 \label{fig:hrd_mm}
  \end{figure} 
  
Simulation work in the context of single-dish IM with MeerKAT further investigated the impact of using different observing strategies and map-making algorithms on the reduction of $1/f$ noise. \citet{irfan24} measured $\alpha$, $\beta$ and $f_{k}$ parameters from survey data and found the raw $1/f$ noise to be between 10 and 20 times higher than the instrumental noise level and partially correlated over frequency. This partial correlation is what enables successful cleaning (40 per cent reduction in $1/f$ noise power) through the removal of 3 PCA modes. Having experimented with both a random scan strategy and the Horizontal Raster Drift (HRD), as well as map-making assuming white noise properties versus map-making including a noise correlation matrix, they found the HRD strategy optimal for minimizing time-correlated noise and that noise-informed map-making reduced $1/f$ contamination by up to 30 per cent. \autoref{fig:hrd_mm} shows the spherically averaged power spectra of simulated $1/f$ noise above the simulated instrumental noise level. The HRD strategy can be seen to reduce $1/f$ noise compared to the random strategy over all scales, except at around 0.04 Mpc$^{-1}$. Map-making with a noise covariance matrix (`MM') can also be seen to bring down the $1/f$ contamination compared to an assumption an Gaussian noise (`Binned').          

%----------------------------
\subsection{Joint inference of $1/f$ noise, gain, and sky signal}
\label{sec: 1/f and map making}
%----------------------------

\citet{zzhang_2025} presents a unified approach that jointly estimates the sky map, the smooth gain components, and the $1/f$ noise parameters, based on a physically motivated data model and performed via high-dimensional Bayesian inference. In this framework, the time-ordered data (TOD) from an autocorrelation detector is modeled as a multiplicative expression of the system temperature, the gain (both smooth and stochastic components), and the radiometer noise. This formulation reflects the radiometer equation in its most general form and naturally encodes the multiplicative coupling between gain, system temperature and noise.
By sampling their joint posterior distribution, the method allows information to propagate throughout the inference chain. In particular, the use of time-domain modeling for $1/f$ gain fluctuations avoids the spurious long-range temporal correlations assumed by conventional DFT diagonal $1/f$ models.

To disentangle smooth gain variations from $1/f$ noise, we model the former as a low-order polynomial and the latter as a stationary Gaussian process with a power-law spectrum. The parameter blocks - sky, gain, and noise - are sampled iteratively within a Gibbs framework. Both sky and gain parameters are updated via iterative generalized least squares, which incorporates an effective additive noise estimate. Crucially, the noise amplitude is not fixed or separately parameterized, but intrinsically coupled to the system temperature through gain amplification, ensuring consistent uncertainty propagation.
For the efficient sampling of $1/f$ noise parameters, a Levinson-recursive algorithm is employed, which exploits the Toeplitz structure of the time-domain covariance matrix. This reduces computational complexity from $\mathcal{O}(N^3)$ to $\mathcal{O}(N^2)$.
Using \textsc{MomentEmu} (see \citealt{zzhang_2025_2}) further speeds up the calculation of the noise covariance matrix, enabling good scalability to SKA-scale datasets.
The combination of Gibbs sampling techniques and efficient linear solvers for each component ensures computational tractability despite the high dimensionality of the joint inference problem.

\begin{figure}
  \centering
  
  % First row

  \begin{subfigure}[b]{\linewidth}
    \includegraphics[width=\linewidth]{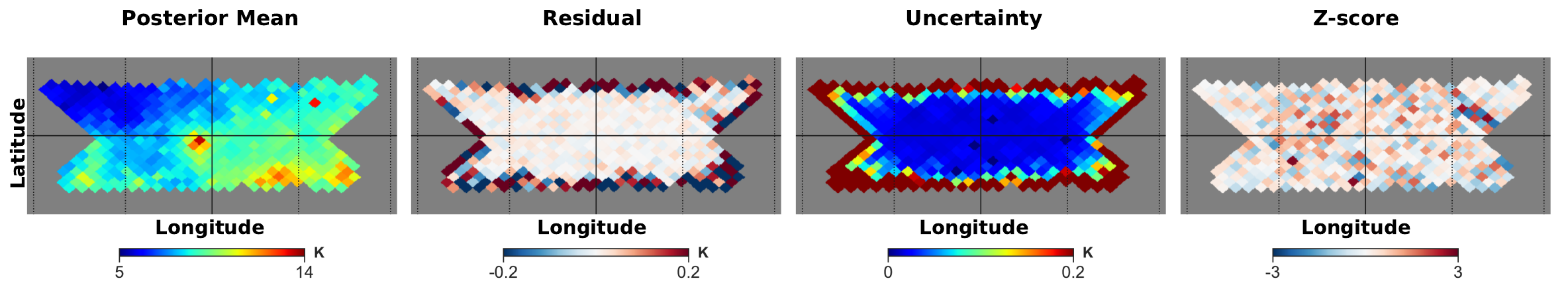}
    \caption{Map reconstructed using the full Bayesian approach. }
    \label{fig: bayesian map-making}
  \end{subfigure}
  \vspace{1em}
  \begin{subfigure}[b]{\linewidth}
    \includegraphics[width=\linewidth]{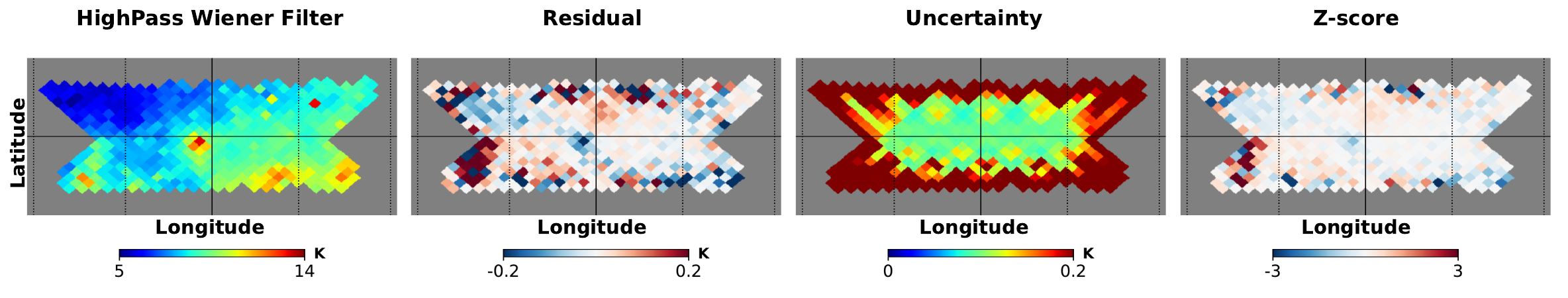}
    \caption{Map reconstructed using the high-pass + Wiener filter method. }
    \label{fig: hpf_wf_map}
  \end{subfigure}
  
  \caption{Comparison between the full Bayesian approach and a conventional `high-pass filter + Wiener filter' method.       
  \textbf{Residual map}:
      Map showing the residual difference between the estimated sky signal  and the true sky input used in simulations. 
  \textbf{Uncertainty map}: 
      Posterior standard deviation of sky parameters marginalized over other parameters, calculated as the sample standard deviation of the draws.
  \textbf{Z-score map}:  defined as the residual map divided by the posterior standard deviation.
  Figure from \citep{zzhang_2025}.}
  \label{fig: map-making comparison}
\end{figure}
To demonstrate the advantages of the full Bayesian approach, \autoref{fig: map-making comparison} compares it with a simple, conventional method that uses a `high-pass filter + Wiener filter' scheme. It can be seen that the traditional method produces larger residuals, particularly within the large-scale structures of the central survey region, and a noisier uncertainty map. Within the survey area, the Bayesian method consistently demonstrates low bias and reduced uncertainty.

\section{Discussion and outlook for the SKA-Mid era}\label{sec:moresys}

In this chapter, we have presented a brief summary of recent advances in component separation and map-making for intensity mapping.
%in the presence of realistic systematic effects. 
The novel methodologies are mostly presented in the context of simulations where it is possible to mimic and isolate specific observational effects. The discussion was necessarily selective, and readers seeking further details are encouraged to consult the references provided.

We have given a large space to blind and semi-blind methods that have been the most successful on data in the last decades (see for example \citet{Cunnington01.2026.SKA,Elahi01.2026.SKA}). This trend highlights a difference with the CMB community, where also non-blind and parametric foreground subtraction have always been considered valid and reliable for the extraction of the cosmological information. The reason for such a difference is easily found: the difference in amplitude between the foregrounds and the signal, and the structure of the signal in frequency space, falsifies some of the classical assumptions of component separation in CMB. This trend could change in the next years due to the signal-to-noise improvement expected in the SKAO era.

Whatever the trend may be, the control of subtle instrumental systematics and residual errors from the calibration pipelines will be critical to the success of future \hi\ IM survey and key to corroborate the solidity of the achievements of IM as a cosmological probe. In conclusion of this chapter, we present a list of the possible challenges left to address.

\emph{Ground pickup}. Although not an instrumental systematic per se, it does contribute to the temperature budget of single dish experiments (as much as 5K for SKA-Mid Band 1). 
Its frequency dependence should be smooth; thus, it could be treated as another foreground and can be mitigated keeping the elevation constant.
However, its convolution with the beam can create a spectral structure. 

\emph{Non-linearities}. Strong RFI can drive the system non-linear, translating into extra gain fluctuations in time for the autocorrelation signal. If we cannot calibrate fast enough in time, this will generate extra fluctuations in time that will make it harder to remove components such as ground pick up and receiver temperature which do show non-trivial frequency dependences.

\emph{Correlated noise}. As discussed, $1/f$ can be problematic for single dish experiments. Experience with MeerKAT has shown this to be small for scales of several minutes and also reasonably smooth in frequency, but it will be critical to also check this stability of the SKA-Mid Band 1 receivers.

\emph{Frequency bandpass errors}. Bandpasses can be well calibrated using good sky models (such as known point sources) but possible changes with time might add frequency fluctuations depending on how often the calibration is repeated. One source of bandpass fluctuations arises from standing waves originating from reflections in the telescope optics. It is common for the optics to change with observing elevation leading to complications for calibration.

The impact of some of these effects, to which single-dish IM could be particularly sensitive, will also depend on the stability and performance of the SKA-Mid. This makes IM a great candidate for the Science Verification.
The methodologies developed thanks to instrumental knowledge, or to the discovery of new systematics in real data, allow not only to tackle the immediate challenges that SKAO precursors and pathfinders are facing, but also to pave the way for the SKAO era.

\section*{Author List Ordering}
The construction of this chapter was led by the first author. All other authors contributed to the writing of the chapter and are listed alphabetically.

\section*{Acknowledgements}
M.~S. is supported by the French government through the France 2030 investment plan managed by the National Research Agency (ANR), as part of the Initiative of Excellence Université Côte d’Azur under reference number ANR- 15-IDEX-01 and by the French Programme National de Cosmologie et Galaxies (PNCG project CIMES).
JLB acknowledges funding from the project UC-LIME (PID2022-140670NA-I00), financed by MCIN/AEI/ 10.13039/501100011033/FEDER, UE.
IPC is supported by the European Union within the Next Generation EU programme [PNRR-4-2-1.2 project No.\ SOE\textunderscore0000136, RadioGaGa]. 
JF thanks the support of FCT - Fundação para a Ciência e a Tecnologia through national funds by these grants: UID/04434/2025 (DOI 10.54499/UID/04434/2025) and 2023.15069.PEX. JF acknowledges the support from FCT in the form of work through the Scientific Employment Incentive program (reference 2020.02633.CEECIND/CP1631/CT0002)
ZZ acknowledges support from the European Research Council (ERC) under the European Union's Horizon 2020 research and innovation programme (Grant agreement No. 948764) and the RadioForegroundsPlus project HORIZON-CL4-2023-SPACE-01, GA 101135036.
JW acknowledges support from the National Natural Science Foundation of China (NSFC, Grant No.\ 12573112).
DC acknowledges support from grant number 192243 of the Swiss National Science Foundation.
MGS acknowledges support from the South African Radio Astronomy Observatory and National Research Foundation (Grant No.\ 84156)
\bibliographystyle{abbrvnat-maxbibnames4}
\bibliography{chapter}

\end{document}